\documentclass[preprint]{aastex}

\newcommand{\kms}{\mbox{km s$^{-1}$}}

\newcommand{\Msun}{\mbox{\,$M_{\sun}$}}
\newcommand{\Lsun}{\mbox{$L_{\sun}$}}
\newcommand{\um}{\mbox{$\mu$m}} 
\newcommand{\co}[2]{\mbox{CO $J = #1 \rightarrow #2$}}
\newcommand{\tco}[2]{\mbox{$^{13}$CO(#1$\rightarrow$#2)}}
\newcommand{\cs}[2]{\mbox{CS(#1$\rightarrow$#2)}}

\newcommand{\skipthis}[1]{}
\newcommand{\hh}{\mbox{H$_2$}} 
\newcommand{\hii}{H\,{\scshape{\small II}}}

\shortauthors{Guzm{\'a}n et al.}
\shorttitle{Search for ionized jets}
\usepackage{amssymb,amsmath,natbib,graphicx}
\begin{document}


\title{Search for ionized jets towards high-mass young stellar objects}

\author{Andr\'es E. Guzm{\'{a}}n\altaffilmark{1}, Guido Garay\altaffilmark{1},
\and Kate J. Brooks\altaffilmark{2} \and Maxim A. Voronkov\altaffilmark{2}}


\altaffiltext{1}{Departamento de Astronom\'{\i}a, Universidad de Chile, Camino el
  Observatorio 1515, Las Condes, Santiago, Chile}\altaffiltext{2}{
CSIRO Astronomy and Space Science, P.O. Box 76, Epping 1710 NSW,
  Australia}

\begin{abstract}

We are carrying out multi-frequency radio continuum observations, using the
Australia Telescope Compact Array, to systematically search for collimated
ionized jets towards high-mass young stellar objects (HMYSOs).  Here we
report observations at 1.4, 2.4, 4.8 and 8.6 GHz, made with angular
resolutions of about 7\arcsec, 4\arcsec, 2\arcsec, and 1\arcsec,
respectively, towards six objects of a sample of 33 southern HMYSOs thought
to be in very early stages of evolution.  The objects in the sample were
selected from radio and infrared catalogs by having positive radio spectral
indices and being luminous ($L_{\rm bol}>2\times10^4~\Lsun$), but
underluminous in radio emission compared to that expected from its
bolometric luminosity. This criteria makes the radio sources good
candidates for being ionized jets.  As part of this systematic search, two
ionized jets have been discovered: one previously published and the other
reported here.  The rest of the observed candidates correspond to three
hypercompact \hii\ regions and two ultracompact \hii\ regions.  The two
jets discovered are associated with two of the most luminous ($7\times10^4$
and $1.0\times10^5$ \Lsun) HMYSOs known to harbor this type of objects,
showing that the phenomena of collimated ionized winds appears in the
formation process of stars at least up to masses of $\sim 20$ \Msun\ and
provides strong evidence for a disk-mediated accretion scenario for the
formation of high-mass stars.  From the incidence of jets in our sample, we
estimate that the jet phase in high-mass protostars lasts for
$\sim4\times10^4$ yr.

\end{abstract}  

\keywords{ISM: jets and outflows 
--- radio continuum: stars --- stars: formation --- stars: individual (G337.4032$-$00.4037)}

\clearpage

\section{INTRODUCTION}
The determination of whether massive stars ($M>8$\Msun) are formed by an
accretion process similar to that inferred for low-mass stars, with an
accreting disk \citep{Shu1987ARA&A}, or via merging processes
\citep[e.g.,][]{Bonnell1998MNRAS} is one of the main observational
challenges in the field of star formation.  Theoretically, an accretion
disk phase could even be more important for high-mass stars, compared to
the low-mass case, in terms of the fraction of the mass of the central
protostar accreted from such a disk \citep{Kuiper2010ApJ}.

One of the striking features of the disk accretion theory is its connection
with the ejection of a collimated jet.  The consensus is that detecting the
presence of such jet is a sufficient condition to ensure that disk
accretion is taking place.  Whether jets are key in allowing disk material
to lose angular momentum, and fall onto the protostar, is still an open
issue \citep{Livio2009pjc}.  However, the latter idea is suggested by
theoretical work in magneto-hydrodynamic modeling of the disk accretion and
jet ejection process.  In addition, jets form a natural candidate to
explain the momentum and energy source of the commonly observed massive
bipolar molecular outflows \citep[e.g.][]{Beuther2002AA,Kim2006ApJ}.
Collimated jets detected toward O-type forming stars would be strong
evidence favoring a disk-mediated accretion process.  On the other hand, if
they are formed via merging of lower-mass stars then neither accretion
disks nor jets are expected \citep{Bally2005AJ}.  The study of the
incidence of jets in high-mass young stellar objects (HMYSOs) will then aid
in identifying the dominating mechanism of massive star formation.

There exists evidence that stars with masses up to $\sim10$ \Msun\ (or
early B-type star) are formed in a disk mediated accretion scenario
\citep[e.g.,][]{Garay1999PASP,Patel2005Natur,Chini2006ApJ}, and there is
growing observational evidence of a disk-mediated accretion collapse for
even more massive objects \citep{Cesaroni2007PrPl,Kraus2010Natur,
  Preibisch2011AA}. The evidence for jets from massive YSOs is rather
scarce, however.  To date there are only a handful of HMYSOs known to be
associated with highly collimated jets and/or Herbig-Haro (HH) objects
\citep[see][]{Guzman2010ApJ}. Most of them have bolometric luminosities
smaller than $2\times10^4$~\Lsun\ corresponding to that of a B0 zero-age
main sequence (ZAMS) star. There is only one O-type YSO with bolometric
luminosity $L>5\times10^4$~\Lsun\ that is associated with a thermal
collimated jet: G343.1262$-$00.0620 (also IRAS 16547$-$4247,
$L\sim6.2\times10^4$~\Lsun; \citealp{Garay2003ApJ};
\citealp{Brooks2007ApJ}).  It is not clear whether the lack of young
massive stars with spectral types earlier than B0 associated with jets
and/or disks is an intrinsic property of the most massive stars or due to
observational disadvantages.  Massive stars are rarer and their
evolutionary time scales are much shorter than those of low mass stars.
With the currently available small sample, made from a collection of
individual serendipitous studies, it is difficult to characterize the jet
phenomena in the process of high-mass star formation.

The presence of jets can be observationally established from high angular
resolution radio continuum observations either directly through their
elongated structures and distinctive spectral indices
\citep{Reynolds1986ApJ} or indirectly by the detection of phenomena
intimately associated to the jets, such as Herbig-Haro objects or aligned
radio lobes.  In this paper we report the first results of a systematic
search for ionized jets, made using the Australia Telescope Compact Array
(ATCA), towards a sample of HMYSOs thought to harbor this type of objects.
In \S \ref{sec-sousel} we describe the selection criteria adopted to build
a sample of jet candidates selected from two catalogs of radio sources
associated with either HMYSOs or ultracompact \hii\ regions (UC\hii).  In
\S \ref{section-observations} we describe the ATCA observations toward six
of the jet candidates.  and in \S \ref{section-results} we present the
results and physical interpretation of the data obtained toward these six
sources.
 
We discuss in \S \ref{sec-discussion} the search outcome and analyze the
consequences of the observed jet incidence.  In the analysis we include the
results obtained toward one of the jet candidates, G345.4938$+$01.4677
(also IRAS 16532$-$3959).  A detailed account of its characteristics and
physical nature was already presented by \citet{Guzman2010ApJ}.
Summarizing, the radio emission corresponds to a string of radio sources
consisting of a compact, bright central component, and four aligned and
symmetrically located lobes.  The central object corresponds to a ionized
jet and the emission from the lobes arises in shocks resulting from the
interaction of the collimated wind with the surrounding medium.  We stress
that this source, which is currently one of the most massive and luminous
($7.0\times10^4$~\Lsun) HMYSO known to be associated with an ionized jet,
was discovered as part of the systematic search reported in this paper.

Finally, we summarize our work and conclusions in \S \ref{section-summary}.
In the Appendix we report the results of observations obtained toward a
sample of five radio sources which were selected in an early stage of this
study.  Three of them are radio sources associated to young high-mass
stars, but they do not fulfill our selection criteria to be a jet
candidate.

\section{SOURCE SELECTION}\label{sec-sousel}

In this section we describe the steps followed to build our sample of jet
candidates towards high-mass young stellar objects. Three requirements were
imposed in selecting the targets:

\begin{itemize}

\item
{\sl Positive radio continuum spectral indices.}
First, and as expected for thermal jets \citep{Reynolds1986ApJ}, we considered 
radio sources in the literature with positive spectral indices
at radio wavelengths.

\item
{\sl Association with luminous infrared sources.}  Since we are searching
for jets associated with high-mass YSOs, we expect the
driving sources to be luminous and still be enshrouded in a dense cloud of
gas and dust.  Thus, most of their luminosity should be re-emitted in the
infrared and far-infrared (FIR).  A second requisite to be
fulfilled by our targets is that they be associated with luminous IRAS
sources.

\item
{\sl Underluminous radio objects.}  We expect radio jets to be only visible
in a very early stage of evolution prior to the ultracompact \hii\ region
phase.  In such an early stage the UV photons from the central protostar
are likely to be trapped by the infalling gas and therefore quench the
development of an UC\hii\ region \citep{Yorke1986ARA&A,Wolfire1987ApJ}. Thus the third
requisite is that the radio luminosity of the targets be considerable lower
than that predicted from the bolometric luminosity.

\end{itemize}

As a starting point to build up our target list, we considered two catalogs
of radio sources: \citet{Urquhart2007AA} and \citet{Walsh1998MNRAS}.
\citeauthor{Urquhart2007AA}~reported radio continuum observations, at 4.8
and 8.6 GHz using ATCA, toward 826 Red MSX Sources 
\citep[RMS,][]{Urquhart2008ASPC}. RMS sources are HMYSO candidates selected by
their colors \citep{Lumsden2002MNRAS} in the \emph{Midcourse Space
  Experiment} (MSX) and 2-MASS bands \citep{Skrustskie2006AJ}.  We
considered all the radio sources that are within 18\arcsec\ (MSX beam) from
the peak position of the associated MSX source, which amounted to 239
objects.  \citet{Walsh1998MNRAS} reported radio continuum observations at
6.7 and 8.6 GHz, using ATCA, towards IRAS sources with colors of
UC\hii\ regions \citep{Wood1989ApJ} and associated with methanol
masers. They reported emission at 8.6 GHz from 177 ultracompact
\hii\ regions.

To be in accordance with the first requirement, we selected from the above
radio sources those which exhibit positive radio continuum spectral index
between the two observed frequencies. 
We find that approximately a 40\% of the radio sources
from both catalogs fulfill this first requirement.
We considered sources with upper limits (or non-detections) at the lower
frequency. In the case of \citet{Urquhart2007AA} the sensitivity in the two
observed bands is similar, hence a non-detection in the lower frequency
band implies a true positive spectral index. In the case of
\citet{Walsh1998MNRAS}, the observations at 6.7 GHz were made with a
bandwidth of 4 MHz, much smaller than that used for at 8.6 GHz (128 MHz),
and therefore are considerably less sensitive than those at the high
frequency. In this case, a non detection does not necessarily implies a
truly positive spectral index. 

To comply with the second requirement, we then selected from the above
sub-sample of radio sources those that are associated --- angular
separation no larger than 25\arcsec\ --- with an IRAS source.  As distances
we adopted the kinematical distances reported by \citet{Faundez2004AA}
(derived from observations of the \cs{2}{1} transition in
\citealt{Bronfman1996AAS}) or by \citet{Urquhart2007AA13CO,Urquhart2008AA}
(derived from observations of the \tco{2}{1} line). In
\citeauthor{Faundez2004AA}~and in the RMS catalog, the near-far distance
ambiguity for sources located inside the solar circle ($R_\odot=8.5$ kpc)
has been resolved for several sources \citep[see][]{Mottram2011AA,Urquhart2012MNRAS}.
For the sources in which the ambiguity remains, we adopted the \emph{near}
distance.  From the IRAS fluxes and the distance, we estimated the
bolometric luminosity using the expression \citep{Casoli1986AA}
\begin{equation}
\label{eq-LIRAS}
 L_{\it IRAS}= 5.44\,\left(\frac{F_{12}}{0.79}+\frac{F_{25}}{2}+\frac{F_{60}}{3.9}+
\frac{F_{100}}{9.9}\right)\left(\frac{D}{{\rm kpc}}\right)^2 ~\Lsun,
\end{equation}
where $L_{\it IRAS}$ is the IRAS FIR luminosity, $D$ is the distance and
$F_{12}$, $F_{25}$ $F_{60}$, and $F_{100}$ correspond to the IRAS fluxes in
Jy measured at the 12, 25, 60, and 100 \um\ bands, respectively. From the
sources with positive radio spectral index, approximately   37\% 
have molecular line observations and FIR luminosity  $L_{\it IRAS}>2\times10^4\Lsun$.
 
To comply with the third requirement, we computed the monochromatic 
radio luminosity, $4 \pi D^2 F_\nu$,  expected from an homogeneous, 
optically thin \hii\  region that is excited by a single star with a 
bolometric luminosity, $L_{\it IRAS}$, using the expression \citep{Spitzer1998BOOK} 
\begin{equation}\label{eq-radvsNi}
\left(\frac{4 \pi D^2 F_\nu}{\rm Jy~kpc^2}\right)
 =0.12 \left(\frac{N_*}{10^{45}~s^{-1}}\right)
\left(\frac{\nu}{8.6 \text{ GHz}}\right)^{-0.11}
\left(\frac{T}{8000{\rm K}}\right)^{0.38}, 
\end{equation}
where $N_*$ is the rate of ionizing continuum photons emitted by a ZAMS
star with that luminosity taken from standard stellar atmospheres models
\citep[e.g.,][]{Panagia1973AJ}. In Eq.~\eqref{eq-radvsNi} $\nu$ is the
observing frequency, and $T$ is the electron temperature of the ionized
gas.  Then we selected from the above subsample those radio sources for
which the observed radio luminosity is smaller than that predicted from the
latter equation by a factor of at least 10, assuming $T=8000$ K.  Note that
Eq. \eqref{eq-radvsNi} implicitly assumes that all the ionizing photons are
absorbed by the plasma. This third and last requisite reduces the number of
jet candidate HMYSOs approximately in 50\%.

While we argue that the third requirement is needed to select ionized jets,
there are other explanations for the radio under-luminosity: (i) The radio
emission at the observed frequency may arise from an optically thick
\hii\ region, in which case the radio luminosity is not proportional to the
number of ionizing photons; (ii) Dust within the \hii\ region absorbs an
important fraction of the ionizing photons, (iii) The distance to the
source is overestimated, (iv) The FIR luminosity is due to multiple
unresolved stars, and  (v) The ionizing photon flux may not be
related to the star bolometric luminosity through standard stellar
atmospheres models.  For instance, it has been found that high accretion
rates ($\dot{M}\sim10^{-3}-10^{-4}$\Msun\, yr$^{-1}$), implies large radius
and low effective temperatures associated with the accreting protostar
\citep{Hosokawa2009ApJ}. High-mass protostars may produce little if any
ionizing photons in their earliest stages.  This explanation has recently
been considered to account of the lack of ionized gas towards high-mass
protostellar objects in the RMS sample \citep{Mottram2011ApJL}.

Table \ref{tab-candidates} presents our sample of 33 jet candidates
associated with high-mass young stars that fulfill the three main
requisites: 23 from RMS and 10 from \citet{Walsh1998MNRAS}.  Columns
(2) and (3) give the right ascension and declination of the radio source,
respectively; cols.~(4) and (5) the observed flux densities at the low and
high frequencies, respectively. The low frequency corresponds to 4.8 or 6.7
GHz, as explained above.  Column (6) give the spectral index; col.~(7) the
kinematic distance; cols.~(8) and (9) the name and estimated FIR luminosity
of the associated IRAS source, respectively; and col.~(10) the reference.
Interestingly, one of the selected jet candidates corresponds to
G343.1262$-$00.0620, the most luminous source known to be associated with a
jet \citep{Garay2003ApJ,Brooks2003ApJ,Rodriguez2005ApJ}.  The list of 33
jet candidates is an important step for the systematic search for jets
toward HMYSOs, and the analysis of the sub-sample presented in this work
form the basis of the work presented in
\citet{Guzman2011thesis}\footnote{The thesis can be retrieved from
  \texttt{http://www.cybertesis.uchile.cl}}. We expect the  objects 
listed in  Table \ref{tab-candidates} correspond to HMYSOs in their 
earliest phase as radio emitting objects.

\begin{figure}
\centering \includegraphics[angle=-90, width=.7\textwidth]{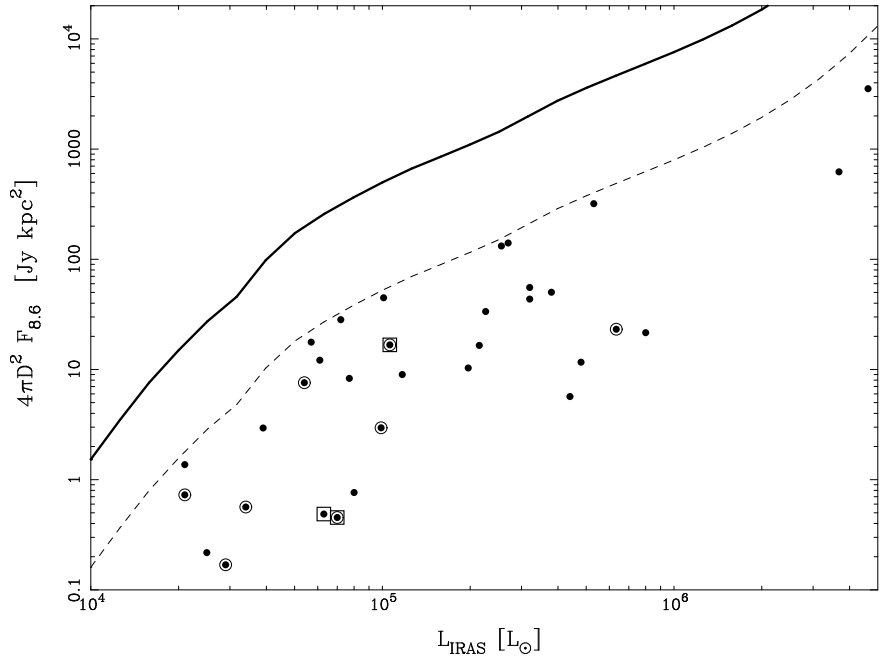}
\figcaption{\baselineskip0.0pt Radio luminosity at 8.6 GHz versus L$_{\it
    IRAS}$ luminosity (Eq.~\ref{eq-LIRAS}) for all jet candidates in Table
  \ref{tab-candidates}.  The continuous line indicates the expected radio
  luminosity of an optically thin \hii\ region ionized by a ZAMS star
  \citep{Panagia1973AJ}.  The dashed line illustrates graphically the third
  selection criterion. Sources observed as part of the jet search described
  in this work are encircled. Squares mark objects G343.1262$-$00.0620
  (also IRAS 16547$-$4247), G345.4938$+$01.4677 (IRAS 16562$-$3959), and
  G337.4032$-$00.4037. \label{fig-LvsQ}}
\end{figure}

Figure \ref{fig-LvsQ} presents a plot of the radio luminosity versus the
bolometric luminosity for all jet candidates.  The
continuous line shows the relation between these quantities for an
optically thin \hii\ region ionized by a ZAMS star \citep{Panagia1973AJ}.
Sources below the dashed line fulfill the third selection criteria.  This
figure clearly illustrates that the jet candidates are underluminous in radio
continuum emission compared to that expected for a uniform optically thin
\hii\ region.

{\section{OBSERVATIONS}\label{section-observations}}

The observations were made using the Australia Telescope Compact Array
(ATCA\footnote{The Australia Telescope Compact Array is funded by the
  Commonwealth of Australia for operation as a National Facility managed by
  CSIRO.}) between 2008 June and 2009 February.  We combined data obtained
using the 1.5B, 1.5C, and 6.0A km array configurations, utilizing all six
antennas and covering east-west baselines from \mbox{30 m} to 5.9 km.
Observations were made at four frequencies: 1.384, 2.368, 4.800, and 8.640
GHz, each with a bandwidth of 128 MHz, full Stokes. Throughout this work we
will refer to these frequencies as 1.4, 2.4, 4.8 and 8.6 GHz,
respectively. The total integration time at each frequency was between 60
and 240 minutes depending on each source, obtained from 10-minute scans
taken over the maximum range of hour angles to provide good (u,v) coverage.
The phase calibrators were observed for 3 min. before and after every
on-source scan in order to correct the amplitude and phase of the
interferometer data for atmospheric and instrumental effects.  The flux
density was calibrated by observing 1934$-$638 (3C84) for which values of
14.95, 11.59, 5.83, and 2.84 Jy were adopted at 1.4, 2.4, 4.8, and 8.6 GHz,
respectively.

Standard calibration and data reduction were performed using MIRIAD
\citep{Sault1995ASPC}.  Whenever the peak emission associated to the source
was above 50 mJy beam$^{-1}$, we applied self-calibration on the phases.
Maps were made by Fourier transformation of the robust-weighted
visibilities (Robust parameter$=0$, see \citealt{Briggs1995PhD}), obtaining
synthesized beams of typical FWHM of $1.6\arcsec\times1.0\arcsec$,
$3.2\arcsec\times1.8\arcsec$, $6.6\arcsec\times3.5\arcsec$, and
10$\arcsec\times$5.0\arcsec\ at 1.4, 2.4, 4.8, and 8.6 GHz, respectively.
The dirty maps obtained were deconvolved using the CLEAN task within
MIRIAD.

The  noise levels attained in the
deconvolved maps at 8.6 GHz were typically  0.1-0.2 mJy beam$^{-1}$, which
is less than that of the RMS and \citet{Walsh1998MNRAS} sample by factors
of about 3 and 5, respectively. Table \ref{tab-obsredpar} gives the phase
tracking center, the on-source total integration times, the FWHM axes of
the synthesized beam obtained at each frequency, and the deconvolved map
noise levels for each observed field. 

{\section{RESULTS}\label{section-results}} 

In this section we present the results of the observations towards six of
the 33 jet candidates presented in Table \ref{tab-candidates}. The observed
parameters of all the radio sources detected in $80\arcsec\times80\arcsec$
fields centered on the target source are given in
Table~\ref{tab-fluxes}. Columns (2) and (3) give the peak position,
cols.~(4-7) the flux densities, and cols.~(8-11) the deconvolved angular
sizes.

The derived parameters, obtained from model fits to the spectra, are
presented in Table \ref{tab-pp}.  Usually the
spectra are well fitted by the theoretical spectrum from an uniform source
of free-free emission, for which we have used the formulae given in
\citet{Garay1993ApJ}.  Columns (2) and (3) gives, respectively, the assumed
distance to the source and the electron temperature used in the fit to the
spectra.  Columns (4) and (5) gives, respectively, the emission measure
(EM) and angular size derived directly from the fit. Column (6) gives the
rate of ionizing photons needed to excite the \hii\ region ($N_i$);
col.~(7) the physical diameter, col.~(8) the mean electron density, and
col.~(9) the type of each radio source.

Additionally, we present  in the Appendix the result of observations made toward five
HMYSOs, selected in an early phase of this work, which are in the radio
catalogs but do not fulfill the final jet candidate selection criteria.

\subsection{Jet candidates}

In this section we present maps of radio continuum emission maps obtained at 
frequencies of 1.4, 2.4, 4.8, and 8.6 GHz, from regions of
80\arcsec$\times80$\arcsec\ in size centered at the position of each jet
candidate observed, indicated by the cross,
as reported in the catalogs of
\citet{Urquhart2007AA} or \citet{Walsh1998MNRAS}.  Contour levels are
in a geometric progression with base between 1 and 2, the value chosen so
each radio map displays the most important resolved and unresolved
features. 
The noise and beam size for each of these maps are listed in
Table \ref{tab-obsredpar}.  Squares in the 1.4 GHz maps indicate the peak
position of the MSX sources in the field. 

In order to better determine the nature  of the radio sources, we
present for each field a three-color mid-infrared (MIR) image made using  8.0,
4.5, and 3.6 \um\ IRAC data from the {\it Spitzer}-GLIMPSE
survey.  The color code for these three filters is red, green, and blue,
respectively.  The brightness scales and color stretches are consistent
throughout this work.  We also present 24 \um\ MIPS images obtained from
the {\it Spitzer}-MIPSGAL survey for the sources G305.7984$-$00.2416 and
G352.5173$-$00.1549.  In addition, in order to support the physical
interpretation developed for G337.4032$-$00.4037, we show a \co{3}{2} line
spectrum obtained from the APEX (Atacama Pathfinder EXperiment) telescope.

\begin{figure}
\plotone{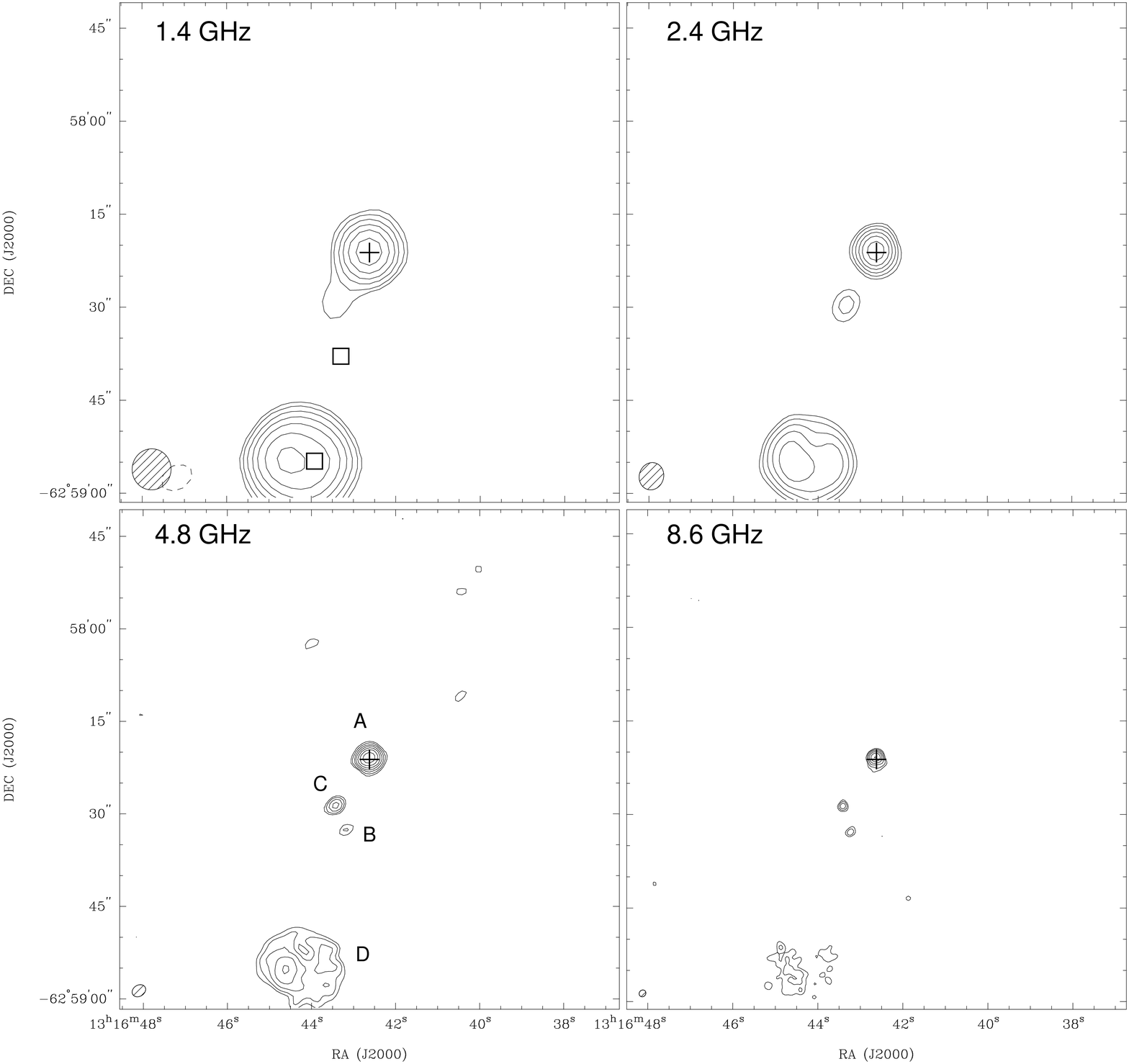} \figcaption {\baselineskip0.0pt ATCA maps of the
  radio continuum emission toward G305.7984$-$00.2416. The cross marks the position
  of the radio source reported by \citet{Walsh1998MNRAS}.  Beams are shown
  in the lower left corner of each panel. 
Top left: 1.4 GHz map. The squares mark the peak position of two MSX 21 \um\ 
sources in the field. Top right: 2.4 GHz map. Bottom left: 4.8 GHz map. 
Labeled are the four radio sources detected in this field. Bottom right: 
8.6 GHz map.
Contour levels are $-$5, 5, 8,
  13, 20, 30, 46, and 70 times 0.16 mJy beam$^{-1}$ for the 1.4 and 2.4 GHz
  maps, and 0.07 mJy beam$^{-1}$ for the 4.8 and 8.6 GHz maps.  
\label{fig-13134}}
\end{figure}

\begin{figure}
\includegraphics[angle=90, width=.7\textwidth]{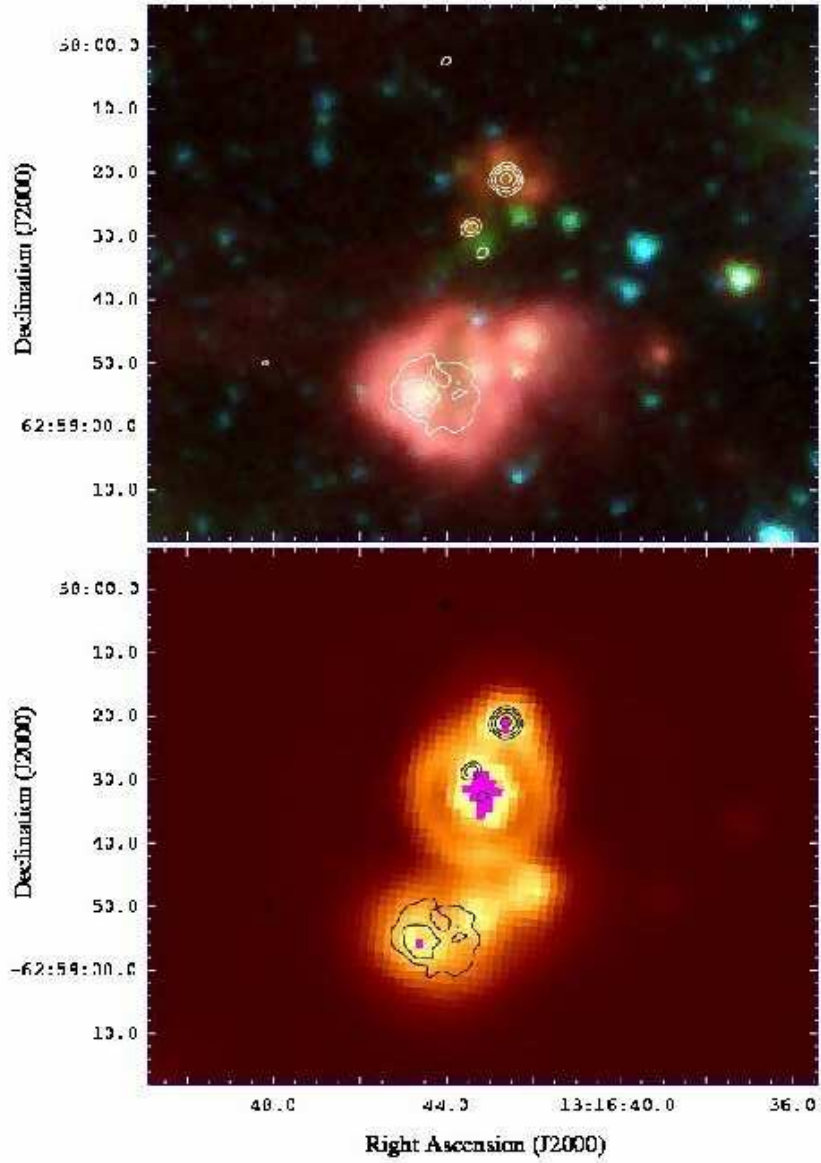} \figcaption {\baselineskip0.0pt 
Comparison between the radio and IR images for
G305.7984$-$00.2416. The white and black contours in each panel show
4.8 GHz data. Top panel: 3 color IRAC image using 8.0, 4.5, and 3.6 \um~
data from {\it Spitzer-}GLIMPSE for red, green, and blue. Bottom panel: 24 \um\ MIPS image from {\it Spitzer-}MIPSGAL. Saturated
pixels are shown in magenta.\label{fig-IR-13134}}
\end{figure}

\begin{figure}
\includegraphics[width=\textwidth]{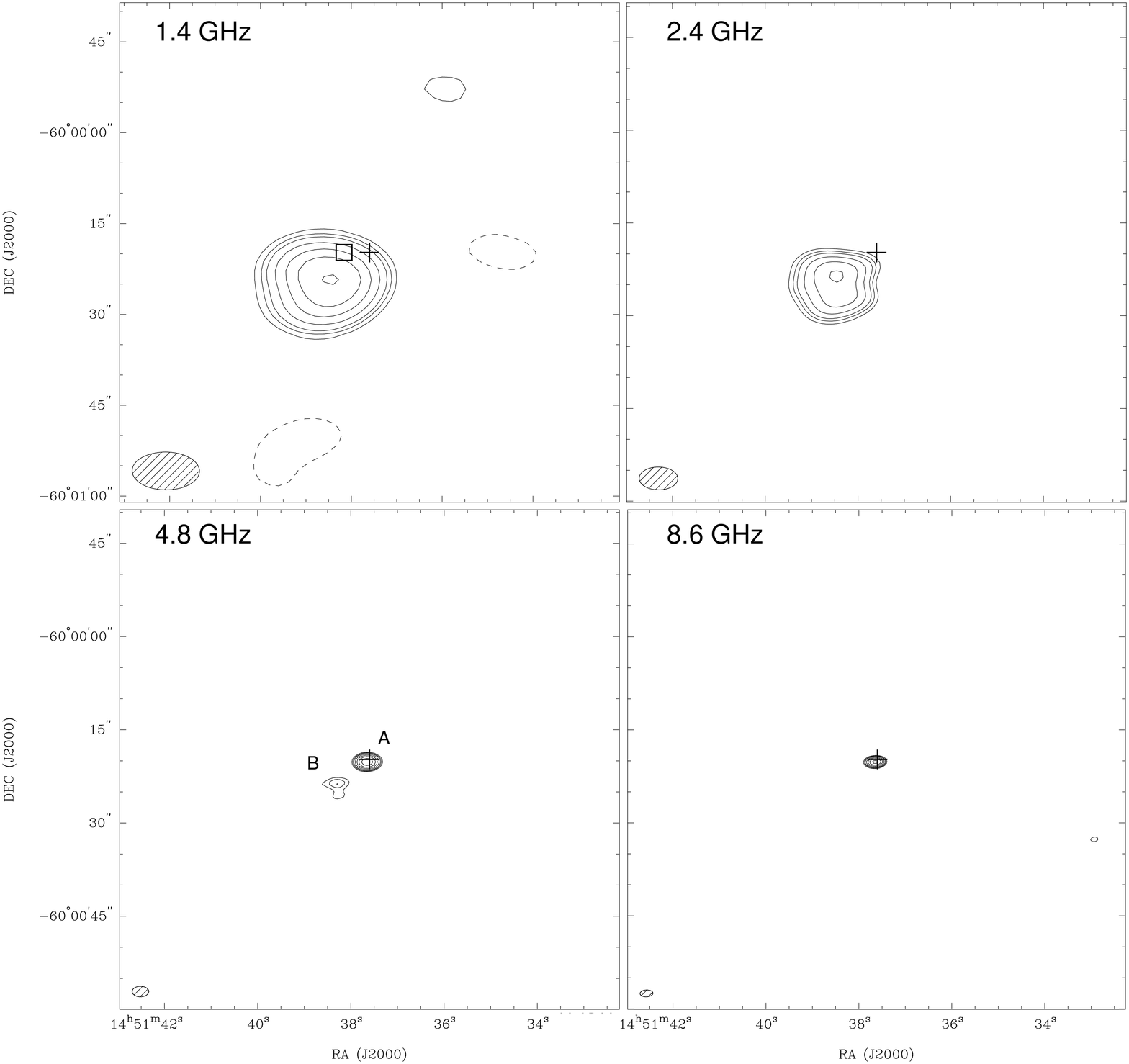} \figcaption {\baselineskip0.0pt ATCA maps of the
  radio continuum emission toward G317.4298$-$00.5612. The cross marks the
  position of the radio source reported by \citet{Urquhart2007AA}.  Beams
  are shown in the lower left corner of each panel.  Top left: 1.4 GHz
  map. The square marks the peak position of the MSX source detected at 21
  \um.  Top right: 2.4 GHz map.  Bottom left: 4.8 GHz
  map. Labeled are the two radio sources detected in this field. 
  Bottom right: 8.6 GHz map.  Contour levels are $-$5, 5, 7, 9, 13, 19,
  27, 40 times 0.5, 0.5, 0.2, and 0.4 mJy beam$^{-1}$ for the 1.4, 2.4, 4.8
  and 8.6 GHz images, respectively.
\label{fig-G317}}
\end{figure}

\begin{figure}
\centering \includegraphics[angle=0, width=\textwidth]{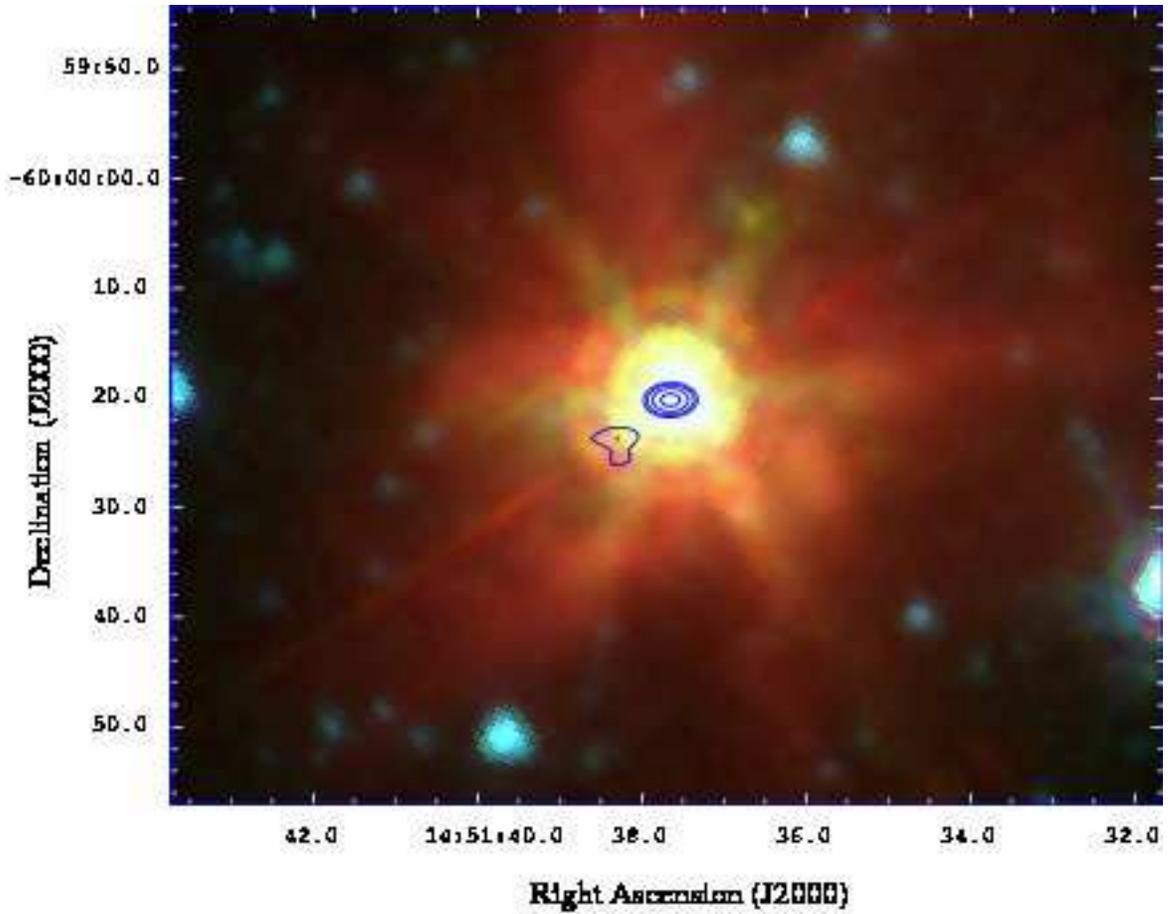}
\figcaption {\baselineskip0.0pt Comparison between mid-infrared and radio
  emission data detected toward G317.4298$-$00.5612.  The color image
  correspond to a three color IRAC image using 8.0, 4.5, and 3.6 \um~data
  from {\it Spitzer-}GLIMPSE for red, green, and blue. Blue contours show 4.8 GHz
  data. \label{fig-IR-G317}}
\end{figure}

\begin{figure}
\includegraphics[width=\textwidth]{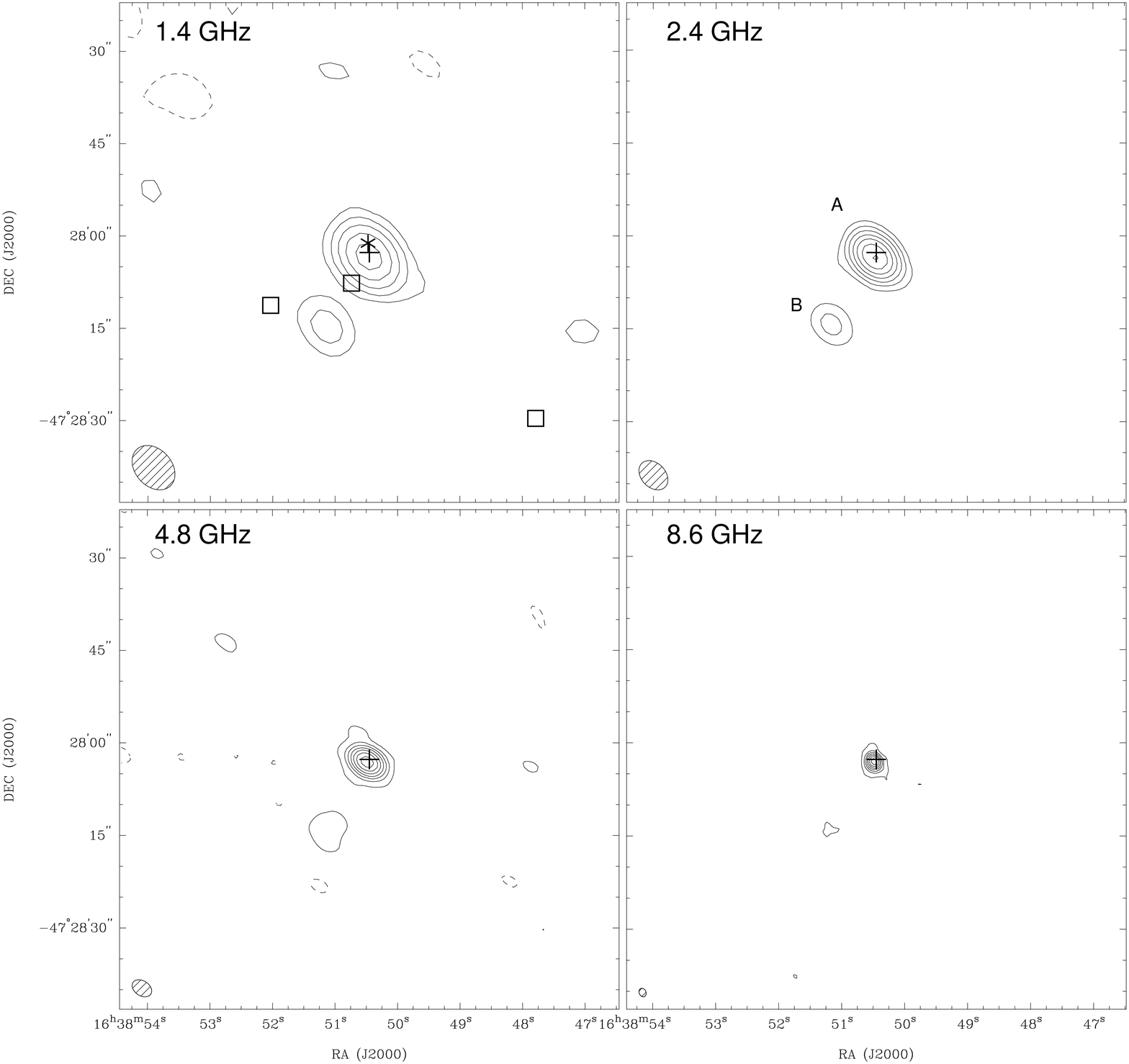} \figcaption {\baselineskip0.0pt ATCA maps of the
  radio continuum emission toward G337.4032$-$00.4037. The cross marks the
  position of the radio source reported by \citet{Urquhart2007AA}.  Beams
  are shown in the lower left corner of each panel.  Top left: 1.4 GHz
  map. The squares show the peak position of the three MSX 21 \um\ sources in
  the field. The star marks the position of the methanol masers detected by
  \citet{Walsh1998MNRAS}. Top right: 2.4 GHz map. Labeled are the two radio
  sources detected in the field. Bottom left: 4.8 GHz map.  Bottom right:
  8.6 GHz map.  Contour levels are $-5$, 5, 14, 28, 50, 81, 129, and 200
  times 0.25 mJy beam$^{-1}$ for the 1.4 and 2.4 GHz images, and $-5$, 5,
  29, 65, 118, 199, 319, and 500 times 0.15 mJy beam$^{-1}$ for the 4.8 and
  8.6 GHz images.
\label{fig-G337}}
\end{figure}

\begin{figure}
\includegraphics[width=\textwidth]{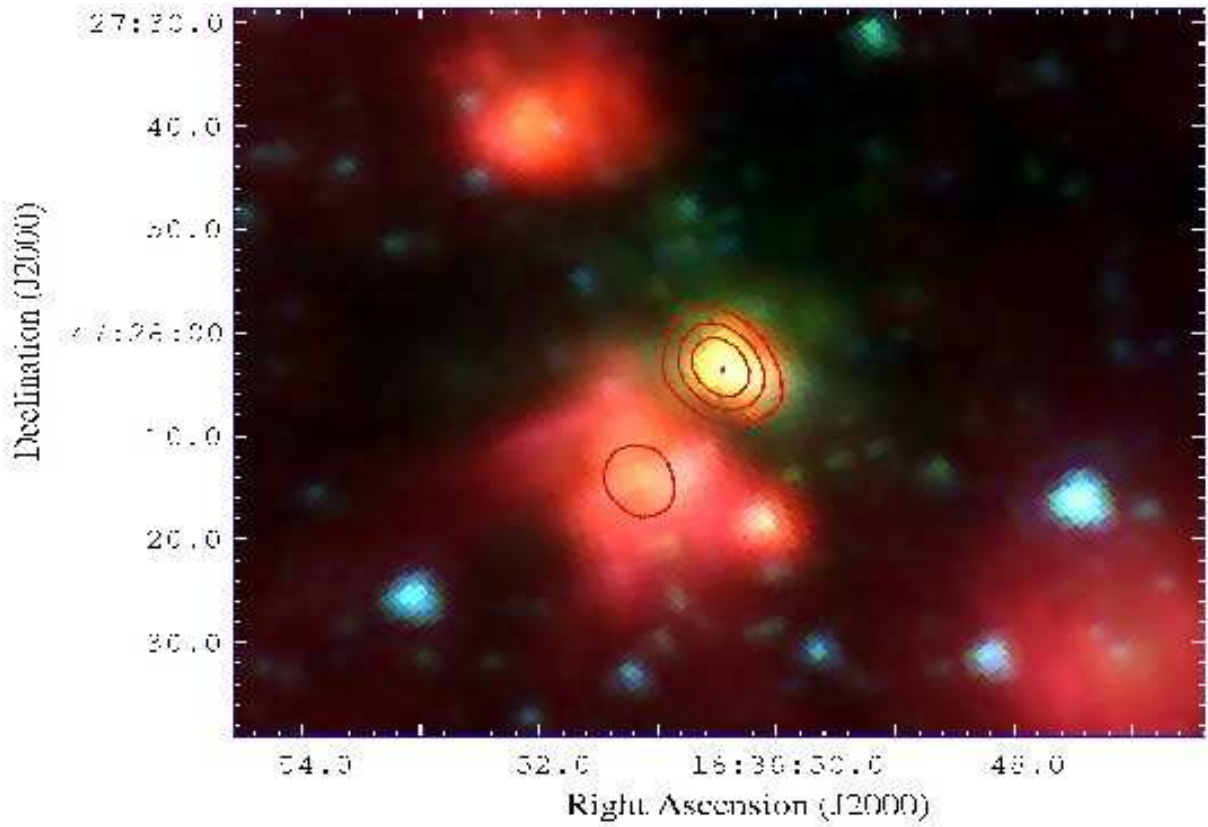} \figcaption
                {\baselineskip0.0pt Three color MIR image toward
                  G337.4032$-$00.4037 made using {\it Spitzer-}GLIMPSE images at 8.0, 4.5, and
                  3.6 \um~for red, green, and blue, respectively. Red
                  contours correspond to the 2.4 GHz radio data.
\label{fig-IR-G337}}
\end{figure}

\begin{figure}\centering
\includegraphics[angle=-90,width=.7\textwidth]{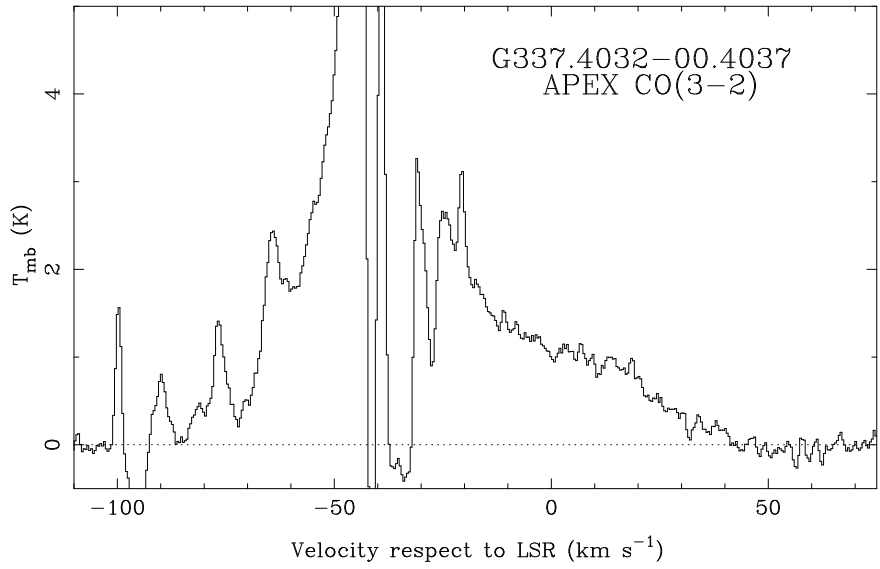}
\figcaption{\baselineskip0.0pt CO(3$\rightarrow$2) spectrum observed toward
  G337.4032$-$00.4037 using APEX. The v$_{\it LSR}$ of the ambient gas associated to the source is
  $-$40.7 \kms.  High-velocity molecular gas emission is detected between $-$85 and
  41 \kms, spanning 126 \kms.
\label{fig-CO32-G337}}
\end{figure}

\begin{figure}
\includegraphics[width=\textwidth]{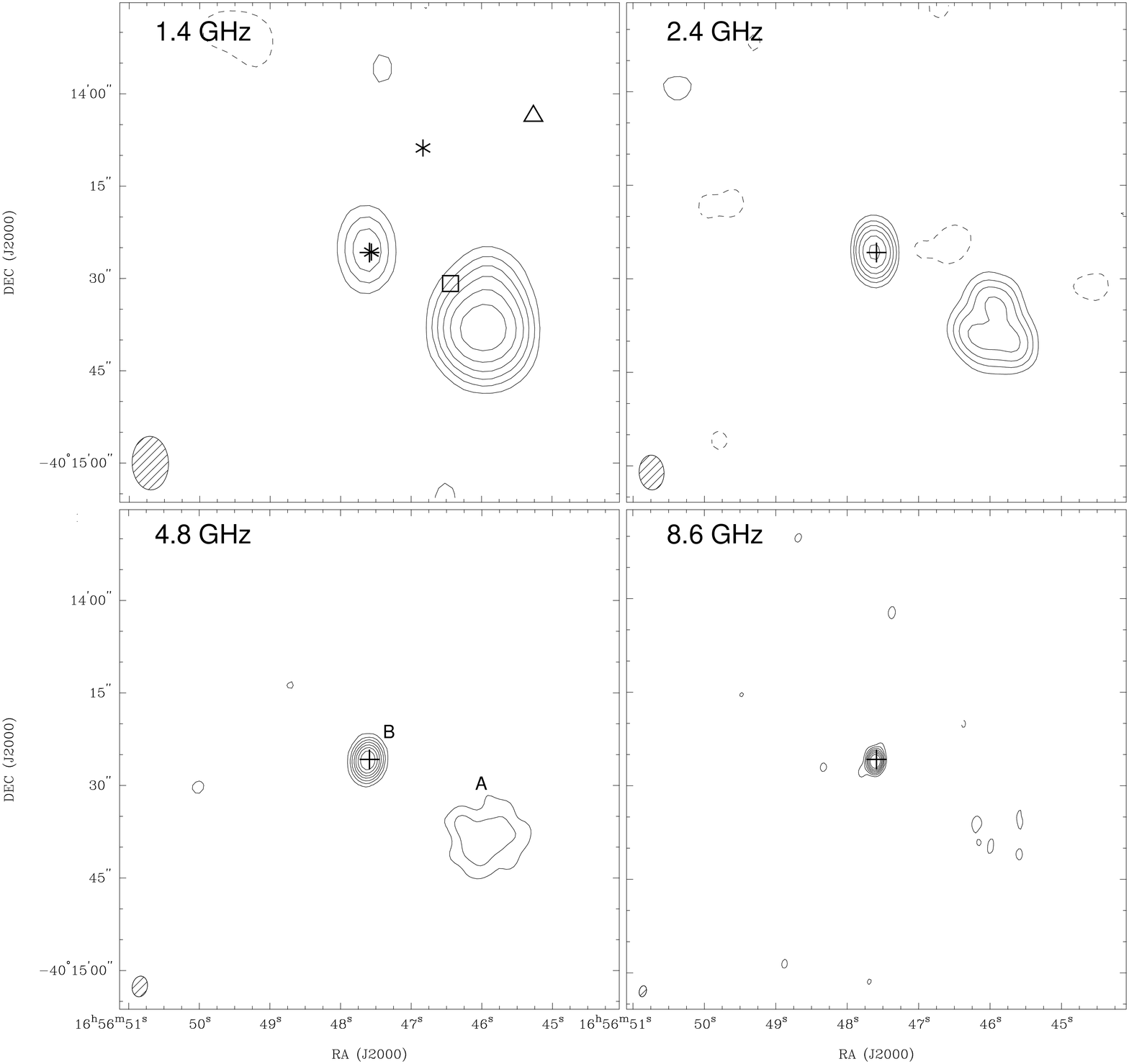} \figcaption {\baselineskip0.0pt ATCA maps of the
  radio continuum emission towards G345.0061$+$01.7944. The cross marks the
  position of the radio source reported by \citet{Urquhart2007AA}.  Beams
  are shown in the lower left corner of each panel.  Top left: 1.4 GHz
  map. The square marks the peak position of the MSX source detected at 21
  \um, the triangle the position of the H$_2$O maser reported by
  \citet{Forster1999A&AS}, and the stars the position of  methanol masers detected by \citet{Walsh1998MNRAS}.
  Top right: 2.4 GHz map. Bottom left: 4.8 GHz map.  Labeled are the two
  radio components detected in the field. Bottom right: 8.6 GHz map.
  Contour levels are $-$5, 5, 11, 19, 31, 50, 78, 120 times 0.4 mJy
  beam$^{-1}$ for the 1.4 and 2.4 GHz images and $-5$, 5, 14, 28, 50, 81,
  129, 200 times 0.4 mJy beam$^{-1}$ for the 2.4 and 8.6 GHz images.
\label{fig-G345.01}}
\end{figure}

\begin{figure}
\includegraphics[width=\textwidth]{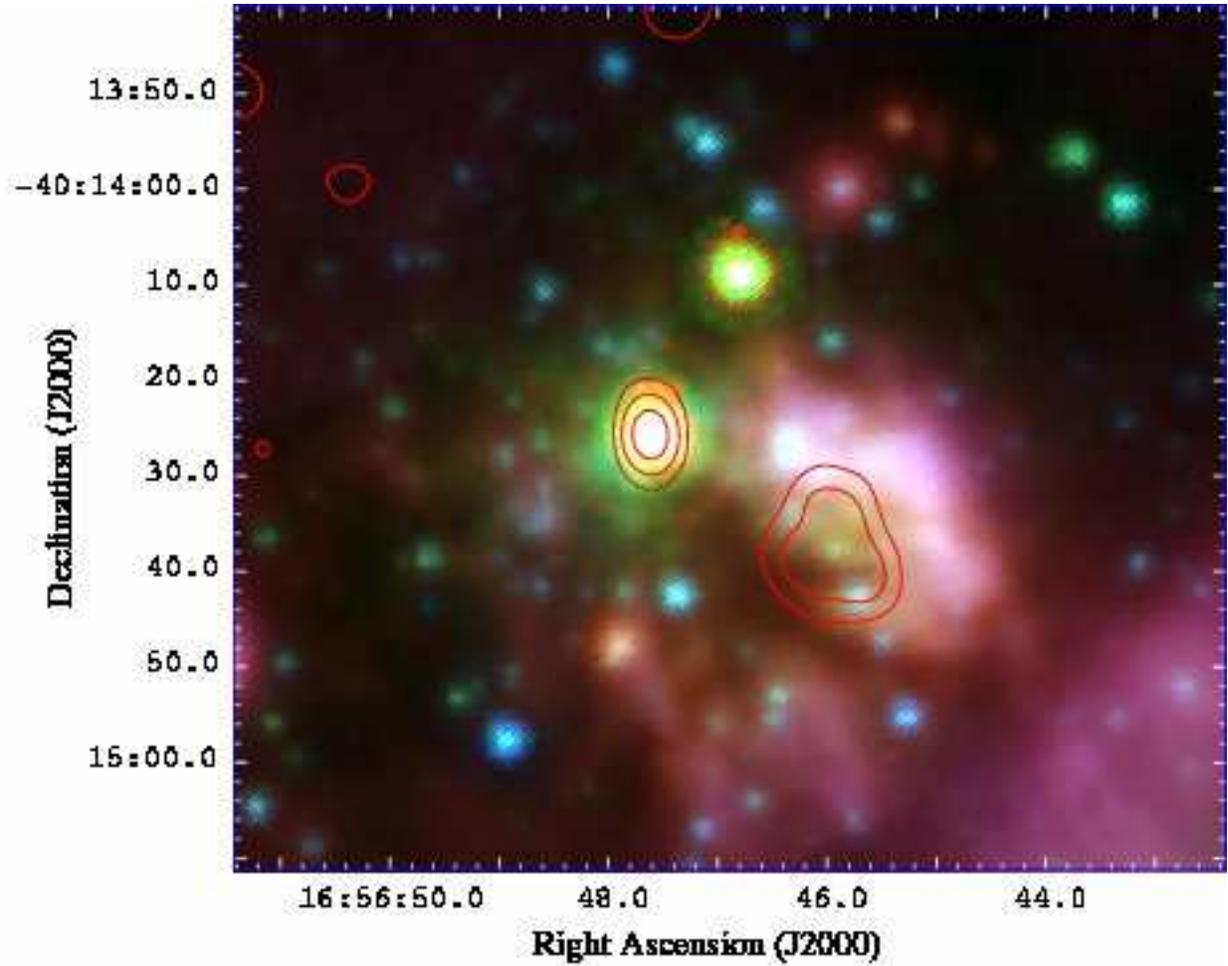} \figcaption {\baselineskip0.0pt 
Three color {\it Spitzer} MIR image of G345.0061$+$01.7944, made using 8.0,
  4.5, and 3.6 \um\ data from {\it Spitzer-}GLIMPSE for red, green, and blue, respectively. Superimposed red contours show the 2.4 GHz radio
  emission.
\label{fig-IR-G345.01}}
\end{figure}

\begin{figure}
\includegraphics[width=\textwidth]{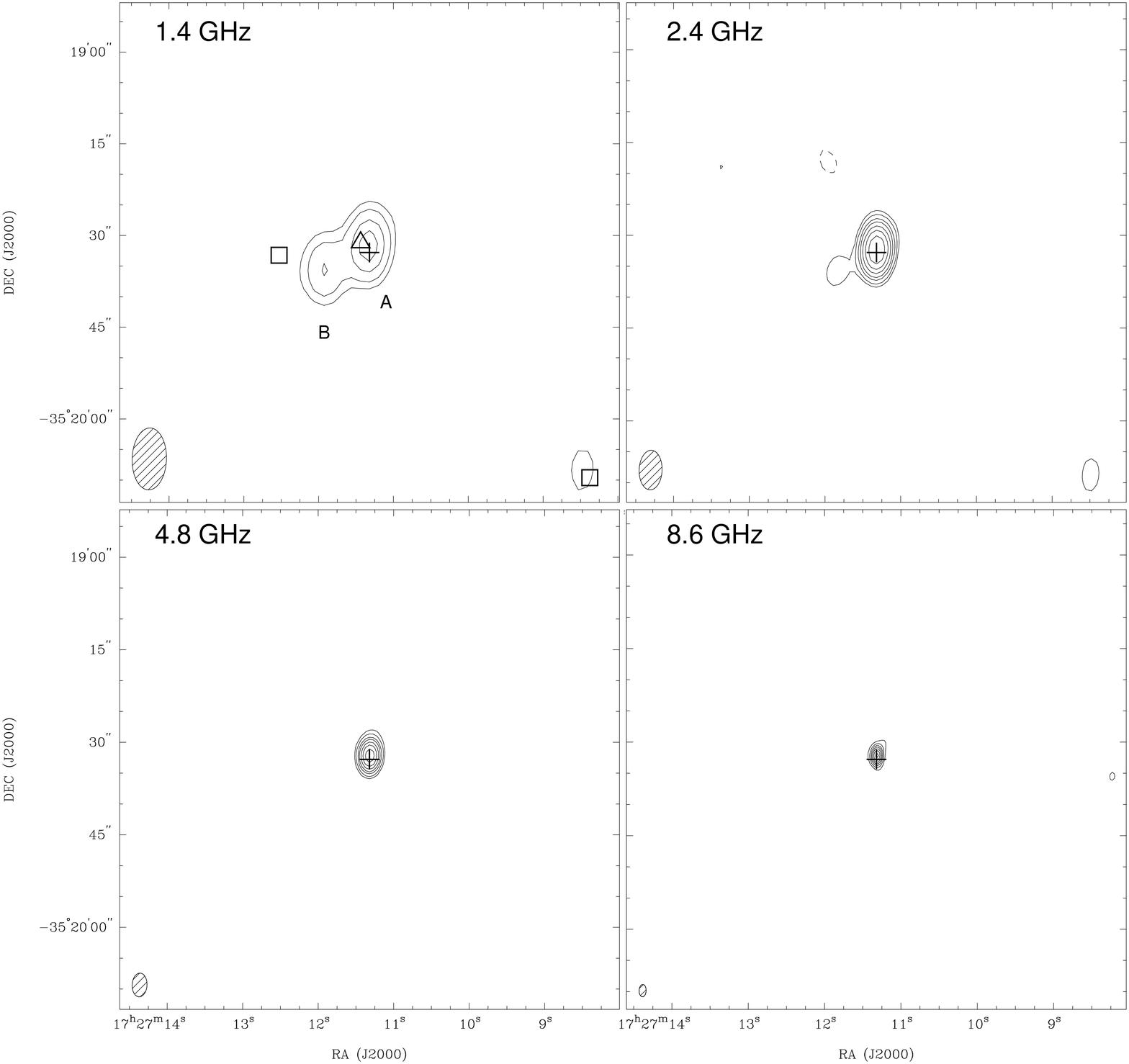} \figcaption {\baselineskip0.0pt ATCA maps of the
  radio continuum emission towards G352.5173$-$00.1549.  Beams are shown in the
  lower left corner of each panel. The cross marks the position of the radio
  source reported by \citet{Walsh1998MNRAS}.  Top left: 1.4 GHz map. Contour levels
  are $-$5, 5, 8, 13, and 18 $\times$ 0.20 mJy beam$^{-1}$. 
  Labeled are the two radio components detected in the field.  Squares mark
  the position of MSX sources. The triangle marks the position of the
  H$_2$O maser detected by \citet{Forster1999A&AS}.  Top right: 2.4 GHz
  map. Contour levels are $-$5, 5, 8, 13, 18, 26, 36, and 50 $\times$ 0.20
  mJy beam$^{-1}$.  Bottom left: 4.8 GHz map. Contour levels are $-$5, 5,
  8, 13, 18, 26, 36, and 50 $\times$ 0.60 mJy beam$^{-1}$.  Bottom right:
  8.6 GHz map. Contour levels are $-$3, 3, 9, 18, 29, 44, 64, and 90
  $\times$ 0.60 mJy beam$^{-1}$.
\label{fig-17238}}
\end{figure}

\begin{figure}
\includegraphics[angle=90,width=.7 \textwidth]{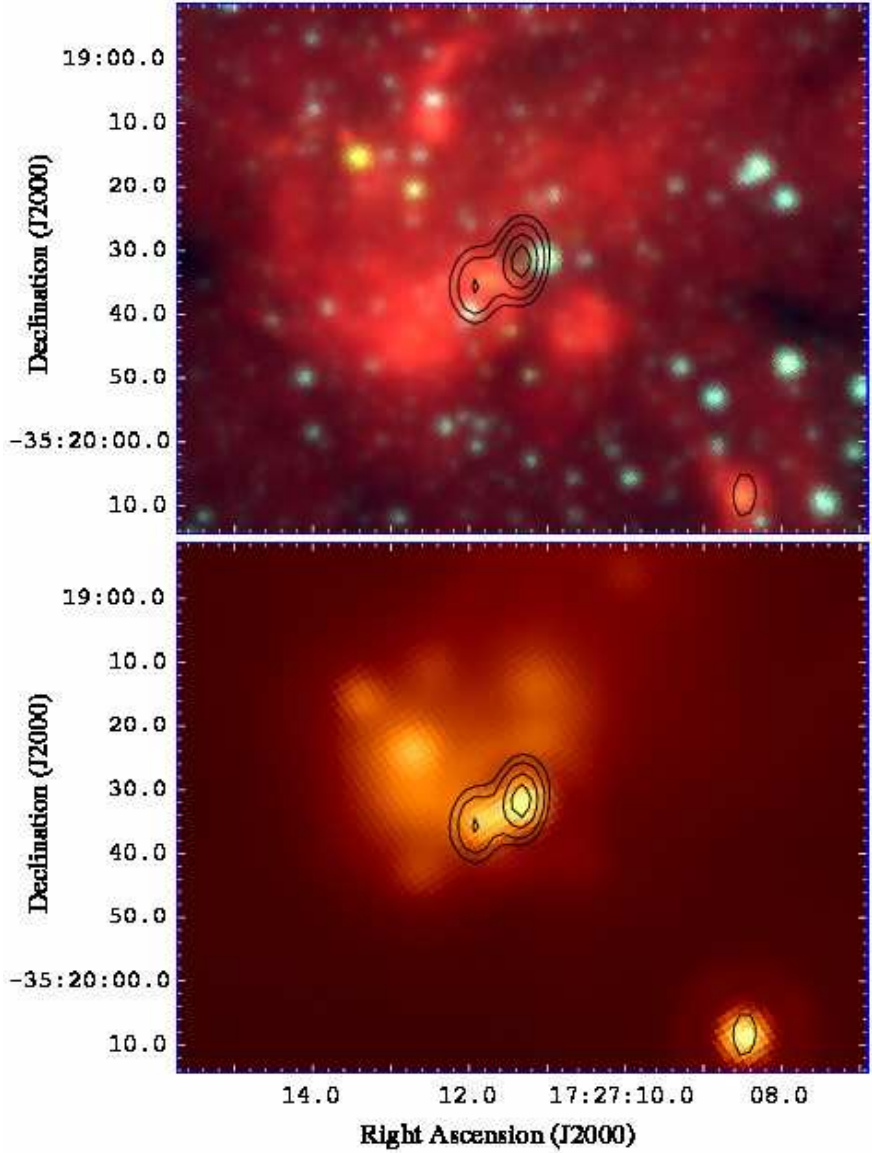} \figcaption
                {\baselineskip0.0pt Mid-infrared images toward
                  G352.5173$-$00.1549, with black contours corresponding to
                  1.4 GHz emission as seen in Fig.~\ref{fig-17238}.
                  \emph{Top Panel:} 3 color IRAC image using 8.0, 4.5, and
                  3.6~\um\ data from {\it Spitzer-}GLIMPSE for red, green, and blue,
                  respectively. The peak of the 1.4 GHz emission is
                  consistent with the peak detected at higher frequencies,
                  and to radio component A. The source seen
                  $\sim3$\arcsec\ to the west is not embedded, and is
                  unrelated with the radio emission. Radio component B
                  appears to be related with the 8.0 \um\ diffuse emission
                  seen at south-east from the peak. \emph{Bottom Panel:}
                  MIPS 24~\um\ emission. Component A is consistent with the
                  peak at MIR wavelengths, and component B also shows MIR
                  counterpart.
\label{fig-IR-17238}}
\end{figure}

\begin{figure}
\includegraphics[width=\textwidth]{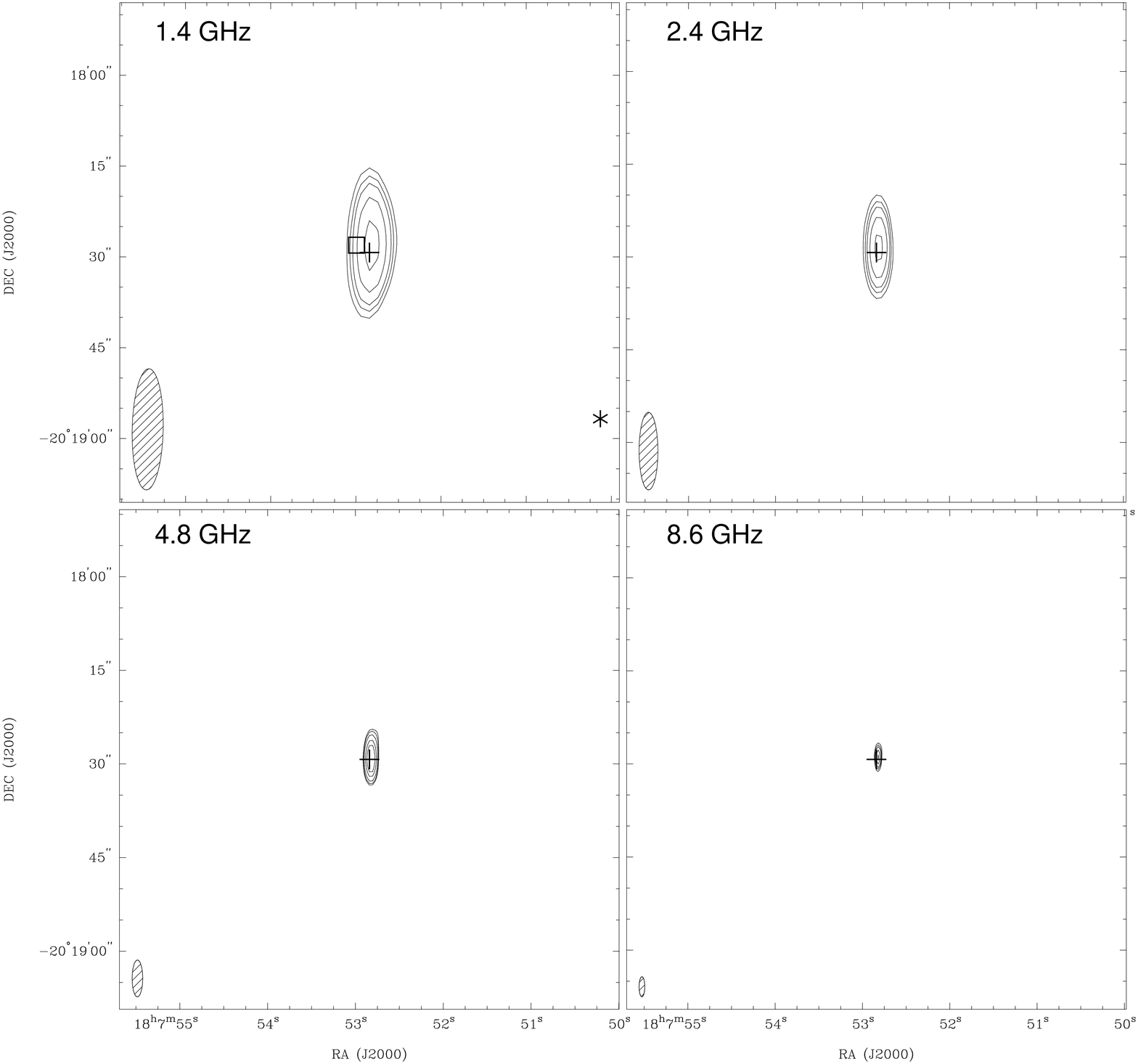} \figcaption {\baselineskip0.0pt ATCA maps of the
  radio continuum emission toward G009.9937$-$00.0299.  The cross marks the
  position of the radio source reported by \citet{Walsh1998MNRAS}.  Beams
  are shown in the lower left corner of each panel.  Top left: 1.4 GHz
  map. The square marks the peak position of the associated MSX source
  G009.9983$-$00.0334 and the star the position of the methanol masers
  detected by \citet{Walsh1998MNRAS}.  Top right: 2.4 GHz map.  Bottom
  left: 4.8 GHz map.  Bottom right: 8.6 GHz map.  Contour levels are $-$3,
  3, 4, 5, 7, and 10 times 0.30 mJy beam$^{-1}$ for the 1.4 and 2.4 GHz
  maps, and 0.23 mJy beam$^{-1}$ for the 4.8 and 8.6 GHz maps.
\label{fig-18048}}
\end{figure}

\begin{figure}
\centering
\includegraphics[angle=0, width=\textwidth]{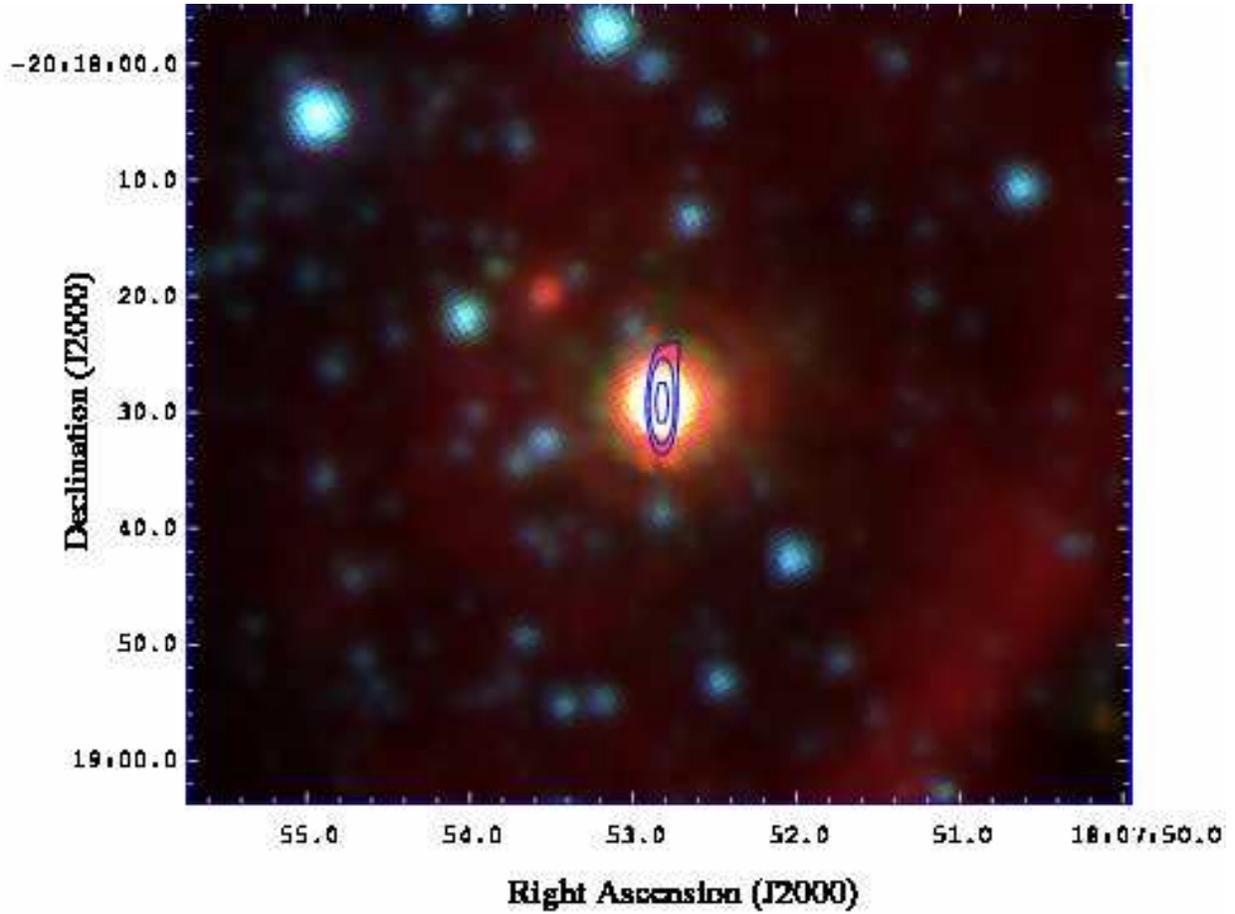} \figcaption
                {\baselineskip0.0pt Comparison between the mid-infrared and
                  radio emission detected toward G009.9937$-$00.0299.  The
                  color image correspond to a three color IRAC image using
                  8.0, 4.5, and 3.6 \um~data from {\it Spitzer-}GLIMPSE for red, green,
                  and blue. Blue contours show 4.8 GHz
                  data. \label{fig-IR-18048}}
\end{figure}

\begin{figure}
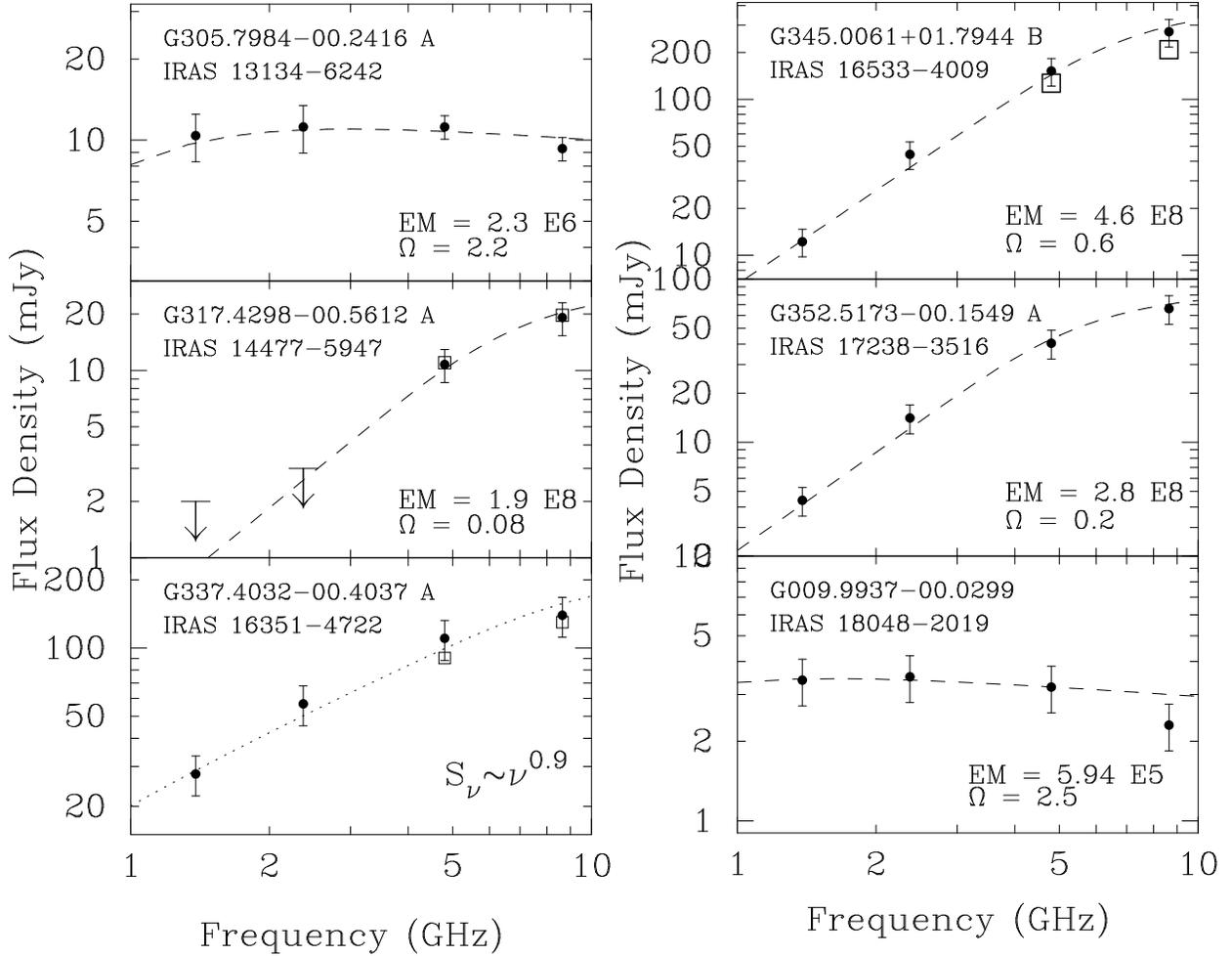

\includegraphics[angle=-90,width=.5
  \textwidth]{fig15-a.eps}\includegraphics[angle=-90,width=.5\textwidth]{fig15-b.eps}
\figcaption {\baselineskip0.0pt Radio continuum spectra of the jet
  candidates observed in this work.  The source name is given in the upper
  left of each panel.  The dashed line corresponds to the best-fit obtained
  with an homogeneous \hii\ region model. The derived emission measure (in
  pc cm$^{-6}$) and solid angle (in arcsec$^2$) are indicated in the lower
  right of each panel.  Dotted lines in the G337.4032$-$00.4037 A panel
  show the best fit obtained with ionized jet models \citep{Reynolds1986ApJ}.
\label{fig-especjetcands}}
\end{figure}

\begin{figure}
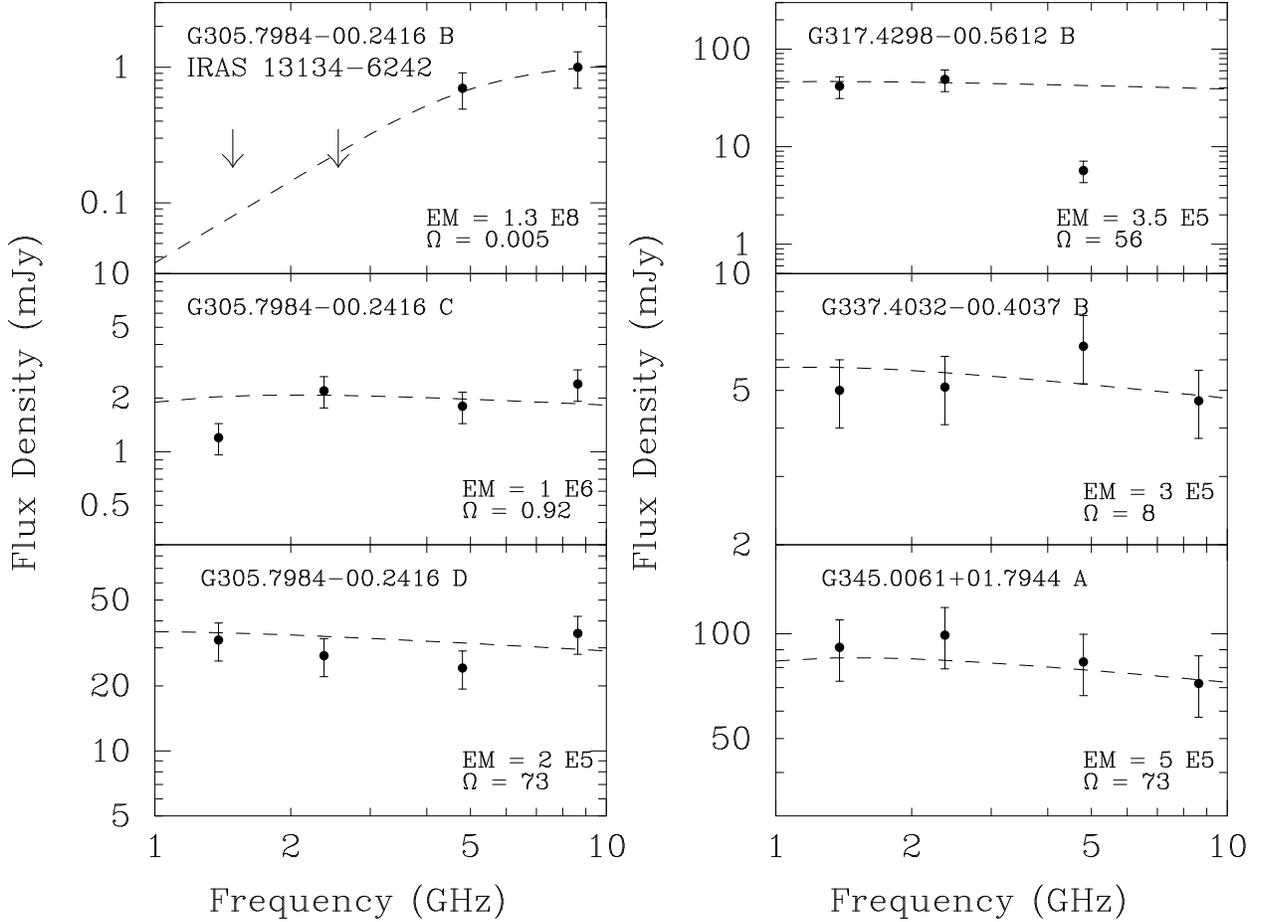

\centering\includegraphics[angle=-90,width=.5\textwidth]{fig16-a.eps}\includegraphics[angle=-90,width=.5\textwidth]{fig16-b.eps}
\figcaption{\baselineskip0.0pt \emph{Left column:} Spectral energy
  distributions of G305.7984$-$00.2416 B, C and D.  \emph{Right column:}
  Spectral energy distributions of G317.4298$-$00.5612 A,
  G337.4032$-$00.4037 B, and G345.0061$+$01.7944 A.  The dashed line shows
  the best-fit obtained with an homogeneous \hii\ region, with the emission
  measure and solid angle indicated in each panel in units of pc cm$^{-6}$
  and arcsec$^2$, respectively.
\label{fig-especjetcands-secondarysources}}
\end{figure}

Figures \ref{fig-13134} to \ref{fig-IR-18048} include the radio maps, and
all mid-infrared supporting figures.  Radio continuum spectra of the jet candidates are
presented in Fig.~\ref{fig-especjetcands}.  Figure \ref{fig-especjetcands-secondarysources}
displays spectra of other sources
detected in the fields.
We also searched
the literature for the presence of water (23.2 GHz) and methanol (6.7 GHz)
masers, thought to trace young stellar environments. We used the catalogs of
water masers presented in \citet{Valdettaro2001AA,Urquhart2009AA} and
\citet{Forster1999A&AS}; and the methanol maser catalogs of
\citet{Pestalozzi2005AA} and  \citet{Walsh1997MNRAS,Walsh1998MNRAS}. In the cases when a reliable position can 
be assigned to  maser emission (via interferometer observations) 
we indicate it in the radio maps.

{\it G305.7984$-$00.2416.}-- This jet candidate, taken from the
\citeauthor{Walsh1998MNRAS}~survey, is associated with IRAS 13134$-$6342,
although it is located $\sim 24\arcsec$ north of the peak position of the
IRAS object.  Figure \ref{fig-13134} shows that there are four radio objects within the
$80\arcsec\times80\arcsec$ region: three
compact components (labeled A, B, and C) and an extended source, labeled D,
located about 35\arcsec\ south of component A.  Components A, C and D were
detected at all frequencies, whereas component B was only detected at the
two higher frequencies. Components B and C are unresolved at all
frequencies.

In order to better determine the nature and evolutionary stage of these
radio sources we present in the top panel of Figure \ref{fig-IR-13134} the
three color MIR image, and in the bottom panel an image of
the {\it Spitzer}-MIPSGAL 24 \um\ emission. 
Superimposed in these two images are
contours of the 4.8 GHz emission.  A comparison
between the radio and MIR images shows that radio component A is associated with
an extended MIR feature which is prominent at 8.0 \um, components B and C
are associated with compact MIR objects GLIMPSE G305.7996$-$00.2438 and
G305.7991$-$00.2447, respectively, and component D is associated with
an extended envelope of diffuse emission most prominent at 8.0 \um, thought
to arise from a photon dominated (or photo-dissociation) region \citep[PDR,][]{Heitsch2007ApJ}.

The jet candidate corresponds to component A, which at 8.6 GHz exhibits a
cometary-like structure with a head toward the north and a tail trailing to
the south.  Its radio continuum spectrum (see Figure~\ref{fig-especjetcands}) is
well fitted by that of a uniform density \hii\ region with an emission
measure of $2.3\times10^6$ pc cm$^{-6}$ and an angular size of 1.4\arcsec.
The ionizing photon flux needed to excite this \hii\ region is
$9.6\times10^{45}$ photons s$^{-1}$.  We conclude that component A
corresponds to a cometary UC\hii\ excited by a B1-B0.5 ZAMS star.  Methanol
maser activity has been detected toward this field
\citep{Pestalozzi2005AA}, although it was not possible to determine its
precise location. 

The radio continuum spectra of components B, C and D are shown in
Figure~\ref{fig-especjetcands-secondarysources}. The spectra of source B shows that its flux
density is rising with frequency, suggesting that it corresponds to an
optically thick \hii\ region.  A fit with a uniform density \hii\ region
model gives an emission measure of $1.3\times10^8$ pc cm$^{-6}$ and an
angular size of 0.066\arcsec.  For component C we derived an emission
measure of $1.0\times10^6$ pc cm$^{-6}$ and an angular size of 0.9\arcsec.
The ionizing photon rate needed to excite this \hii\ region is $N_i
\sim2\times10^{45}$~s$^{-1}$, which can be provided by a B1 ZAMS star.  For
component D we derived an emission measure of $2.0\times10^5$ pc cm$^{-6}$
and an angular size of 8.1\arcsec. The ionizing photon rate needed to
excite this extended and diffuse \hii\ region is
$N_i\sim3\times10^{46}$~s$^{-1}$, which can be provided by a B0.5 ZAMS
star.

The G305.7984$-$00.2416 region may be representative of what is likely to
see within the maternities of high-mass stars: a cluster of young stars in
different evolutionary stages.  We find that the target radio source in the
field is not a jet but an UC \hii.  The weakest radio source detected in
the region (Component B) is particularly striking because, as shown in
Fig.~\ref{fig-IR-13134}, it is associated with a green object in the three
color Spitzer image, possibly indicating shock activity.  Extended green
objects \citep[EGOs,][]{Chambers2009ApJS,Cyganowski2008AJ} are thought to
probe shocked gas regions associated with HMYSOs through vibrational
\hh\ and CO bands.  However, hydrogen Br$\alpha$ contribution cannot be
ruled out \citep{Qiu2011ApJ}.  Even though a consistent scale have been
applied to make all three-color IRAC images presented in this work,
precaution is needed in interpreting the images.  The community has not
adopted an standardized procedure to make these images, casting doubts on
the physical significance of some of the ``green fuzzies''
\citep[e.g.][]{DeBuizer2010AJ}.

The position of Component B is consistent with the peak of the emission in
the MSX band E (21.34 \um\,) image (marked with a square in
Fig.~\ref{fig-13134}), and it also seems to be the brightest source in the
24 \um\ MIPS image (Fig.~\ref{fig-IR-13134}), despite the saturation of the
central pixels.  This evidence suggests that component B is actually the
most embedded, and possibly the youngest, object in the region.  Using the
GLIMPSE flux densities and assuming that the MSX E band flux arises
entirely from this source, the YSO spectral energy distribution (SED) model
fitter of \citet{Robitaille2007ApJS} gives a total luminosity of
$3.3\pm0.9\times10^3$~\Lsun, equivalent to that of a B2 ZAMS star. The
ratio between the ionizing photon flux derived from the radio SED fitting
($N_i\sim 1.2\times10^{45}$ s$^{-1}$) and that inferred from the total luminosity 
($N_\star\sim 6.3\times10^{44}$ s$^{-1}$) is 1.9, implying that this radio source is slightly
under-luminous in radio.  Given its characteristics, particularly the
association to green fuzzy emission, we suggest that component B could
represent the transition phase as an early B-type star grows onto an O-type
star. The excess emission at 4.5 \um\ may arise from shocks produced in the
infalling gas.

{\it G317.4298$-$00.5612.}-- This jet candidate is associated with the
RMS/MSX source G317.4298$-$00.5612 (also IRAS 14477$-$5947). The
kinematical distance of the object is $15$ kpc, with no ambiguity
\citep[$v_{LSR}=27.4$~\kms\ in the CS(2$\rightarrow$1)
  line,][]{Bronfman1996AAS} and assuming a flat rotation curve
($\Theta=220$~\kms). 
From the bolometric flux given by
\citet{Mottram2011AA} we estimate a total luminosity of
$L\sim3.2\times10^5$~\Lsun, equivalent to that of a ZAMS star with an
spectral type between O6 and O5.5. Water maser
has been detected toward this source by \citet{Urquhart2009AA}.   

Figure \ref{fig-G317} shows that there are two radio objects within the $80\arcsec\times80\arcsec$ region: a compact component (labeled A),
seen only at 4.8 and 8.6 GHz, and an extended component (labeled B)
mainly seen at 1.2 and 2.4 GHz and completely resolved out at 8.6 GHz.

Component A corresponds to the jet candidate. At 1.4 and 2.4 GHz this
object is immersed within the more extended component so we could not
measure its flux densities. In the frequency range between 4.8 and 8.6 GHz,
the flux density rises with frequency.  Since the source is unresolved,
from these data it is not possible to discern whether this object is a jet
or an optically thick \hii\ region. If an \hii\ region, then a fit to the
radio continuum spectrum with an uniform density model (see
Fig.~\ref{fig-especjetcands}) indicates an emission measure of
$1.9\times10^8$ pc cm$^{-6}$ and an angular size of 0.27\arcsec. The rate
of ionizing photons required to excite this region of ionized gas is
$7.2\times10^{47}$ s$^{-1}$, which can be provided by a O9.5 ZAMS star.

The radio continuum spectrum of component B (see
Figure~\ref{fig-especjetcands-secondarysources}) suggests it corresponds to an extended
optically thin \hii\ region.  A fit with a theoretical spectra of an
homogeneous constant density \hii\ region gives an emission measure of
$3.5\times10^5$ pc cm$^{-6}$ and an angular size of 7.6\arcsec. The rate of
ionizing photons required to excite the \hii\ region is $9.3\times10^{47}$
s$^{-1}$, which can be provided by a O9.5 ZAMS star.

Figure \ref{fig-IR-G317} shows the  three color mid-infrared IRAC image toward this region.
Component A is associated  with a bright MIR
source with colors similar to a PDR associated to an  UC\hii\ region.

The sum of the luminosities of the ZAMS stars exciting components A and B
is $8\times10^4$~\Lsun, about 4 times smaller than the bolometric
luminosity of the whole region. This discrepancy may be solved if component
A is not an \hii\ region but an ionized jet, in which core the exciting source
could have a larger luminosity than that implied from the observed radio
flux density. Further observations are needed to confirm this interpretation.

{\it G337.4032$-$00.4037}.-- This jet candidate is associated with the RMS
source G337.4032$-$00.4037 (also IRAS 16351$-$4722).  Figure
\ref{fig-G337} shows that there are two radio
sources within the $80\arcsec\times80\arcsec$ region: a bright compact component (labeled A) and a weak
component, labeled B, located about 13\arcsec\ southeast of component A.
Component A, which corresponds to the jet candidate, is unresolved at all
frequencies, whereas component B is resolved out at 4.8 and 8.6
GHz. Methanol maser activity has been associated to component A by
\citet{Walsh1998MNRAS}, whereas no water maser was detected by
\citet{Urquhart2009AA} at 0.25 Jy sigma level.

Figure \ref{fig-IR-G337} shows the  three-color IRAC image of the MIR emission 
detected toward this region, and in red  contours  the 2.4 GHz data.
This Figure shows that component A is associated with the central, bright,
greenish-like object seen in the {\it Spitzer} image.
Assuming that this star forming region is at a distance
of 3.2 kpc \citep{Faundez2004AA}, then from the bolometric flux reported by
\citet[][their component ``B'']{Mottram2011AA},
we obtain a bolometric luminosity of $2.7\times10^4$~\Lsun,
equivalent to that of B0 ZAMS star.

The radio continuum spectrum of component A (see Figure~\ref{fig-especjetcands})
shows that the flux density steadily increases with frequency. A power law
fit gives an spectral index of $0.9\pm0.15$, consistent with a pressure
confined jet \citep{Reynolds1986ApJ}.  Using equation (19) of
\citet{Reynolds1986ApJ},  assuming typical values \citep[e.g.][]{Guzman2010ApJ}
 of 45$^\circ$ for the
inclination, 0.2 rad for the jet aperture, a turnover frequency of 20 GHz
and a wind velocity of 300 \kms, we obtain a mass loss rate of
$\dot{M}\approx2\times10^{-5}\Msun$~yr$^{-1}$.  

Our interpretation of this source as a jet is supported by the association
to a very high velocity molecular outflow.  Figure \ref{fig-CO32-G337}
shows the \co{3}{2} emission detected toward G337.4032$-$00.4037 using the
Atacama Pathfinder EXperiment (APEX). A description of the instrument and
its performance is given by \citet{Gusten2006AA} and a detailed account of
these observations can be found in \citet{Guzman2011thesis}.  The presence
of high-velocity gas is evident from Fig.~\ref{fig-CO32-G337}, where the
line emission can be detected in a $\sim120$ \kms\ range. The ambient gas
velocity respect to the local standard of rest (LSR) is $-$40.7 \kms.

Radio continuum spectra similar to the one exhibited by  this source have
usually been interpreted as arising from  hypercompact \hii\ regions \citep[HC\hii Rs,][]{Franco2000ApJ} with power-law dependence of the density with radius
\citep[e.g.,][]{Panagia1975AA,Avalos2006ApJ}.  For an spherical
\hii\ region in which the density follows $n_e\propto r^{-\beta}$, the
flux density should exhibit a power-law dependence with frequency,
$S_{\nu}\propto \nu^{\alpha}$, with $\alpha=(2\beta-3.2)/(\beta-0.5)$.
There is no hint of an spectrum turn over at high
frequencies, indicating that the emission does not reach optically thin
conditions in the whole observed frequency range.  Assuming ionization
equilibrium, with $N_\star\sim2.6\times 10^{47}$ s$^{-1}$(appropriate for a
ZAMS star with $2.7\times10^4$~\Lsun), and that the power-law index of the
density profile is 2.5 (determined from the above relationship), the observed
spectrum and source size can be well reproduced if the region has an inner
radius of 0.0032 pc, an outer radius of 0.044 pc and an electron density of
$3.6\times10^5$ cm$^{-3}$ at the inner radius.  We note, however, that the
jet and the HC\hii R models have many free parameters, and that these
solutions are not unique.

Figure \ref{fig-IR-G337} also shows that component B is associated with the
bright, central part of an extended, diffuse MIR object, most prominent at
8.0 \um\ and  likely to correspond to a PDR around an \hii\ region.
From the bolometric flux reported
by \citet{Mottram2011AA}, we derive a bolometric luminosity for component B
of 8.4$\times 10^{4}$~\Lsun. The flux density expected at 1.4 GHz from 
an optically thin \hii\ region excited by a star with that luminosity is
$\sim0.5$ Jy, considerably larger than the measured 5 mJy. This implies, as
our observations indicate, that the flux density is resolved out by the
interferometer even at 1.4 GHz.

{\it G345.0061$+$01.7944.}-- This jet candidate is associated with the
RMS/MSX source G345.0061$+$01.7944 (also IRAS 16533$-$4009).  Figure
\ref{fig-G345.01} shows that there are two radio objects within the
$80\arcsec\times80\arcsec$ region: a compact component (labeled B) and an
extended source, labeled A, located about 22\arcsec\ southwest of component
B.  Component B, which corresponds to the jet candidate, is unresolved at
all frequencies whereas component A is partially resolved out at 4.8 GHz
and highly resolved out at 8.6 GHz.

Figure~\ref{fig-IR-G345.01} presents the three-color image of the MIR
emission toward this region, together with red contours displaying the 2.4
GHz data.  This Figure shows that component B is associated with a bright,
compact MIR source with enhanced 4.5 \um~band emission (green color) whereas
component A is associated an extended source of diffuse emission most
prominent at 8.0 \um.  Methanol maser activity has been detected by
\citet{Walsh1998MNRAS} toward two positions: one associated to radio component
B, and another located approximately 20\arcsec\ to the northwest. These
positions correspond to the position of the ``green'' sources seen in
Fig.~\ref{fig-IR-G345.01}.  Water maser activity has been detected in the
field by \citet{Forster1999A&AS}, but not directly associated to any radio or
IR source.
  
The radio continuum spectrum of component B (see Figure~\ref{fig-especjetcands})
suggests it corresponds to an optically thick \hii\ region.  A fit with a
theoretical spectra of an homogeneous constant density \hii\ region, gives
an emission measure of $4.6\times10^8$ pc cm$^{-6}$ and an angular size of
0.73\arcsec. The rate of ionizing photons required to excite the
\hii\ region is $1.7\times10^{47}$ s$^{-1}$, which can be provided by a B0
ZAMS star.  Component B is associated with one of the three IR sources
reported by \citet{Mottram2011AA} within the region (their component
``C''). From its bolometric flux, of $1.19\times 10^{-7}$ erg s$^{-1}$
cm$^{-2}$, and assuming it is at the near distance of 1.7 kpc, we derive
that this IR source has a luminosity of $L_{bol}=1.1\times10^{4}$~\Lsun~,
equivalent to that of a B0.5 ZAMS star. The ionizing photon flux provided
by such a star ($N_\star\approx1.7\times10^{46}$ s$^{-1}$) is about an order of
magnitude smaller than that implied from the modeling of the radio spectra,
being this source \emph{overluminous} in radio rather than underluminous.
We note however that there is considerable uncertainty in the determination
of the bolometric flux: taking a value closer to the  upper limit of the 
uncertainty range given by \citeauthor{Mottram2011AA}~($1.94\times 10^{-7}$ erg s$^{-1}$ cm$^{-2}$), 
 makes the radio and IR luminosities consistent with a photoionized region.

The radio continuum spectrum of component A indicates that it corresponds
to an optically thin region of ionized gas. A fit with a theoretical
spectra of an homogeneous constant density \hii\ region, gives an emission
measure of $5.0\times10^5$ pc cm$^{-6}$ and an angular size of
8.0\arcsec. The rate of ionizing photons required to excite this
\hii\ region is $2.2\times10^{46}$ s$^{-1}$, which can be provided by a
B0.5 ZAMS star.  Two of the three MIR sources identified by
\citet[][their components A and B]{Mottram2011AA}  are associated with this
radio component, but none is located at the peak of the radio emission.

{\it G352.5173$-$00.1549.}-- This jet candidate, taken from the
\citet{Walsh1998MNRAS} survey, is associated with IRAS 17238$-$3516. Figure
\ref{fig-17238} shows that there are two radio sources within the
$80\arcsec\times80\arcsec$ region: a bright compact component (labeled A)
detected at the four frequencies, and a weaker, presumably extended
component (labeled B) only seen at 1.4 and 2.4 GHz and resolved out at the
higher frequencies.

Component A, which corresponds to the jet candidate, can be well modeled by
a homogeneous \hii\ region with $T_e \sim 15000$ K, $\text{EM} \sim 2.8\times10^8$
pc cm$^{-6}$ and an angular size of 0.42\arcsec\ (Figure
\ref{fig-especjetcands}).  The rate of ionizing photons required to excite
this \hii\ region is $3.2\times10^{47}$ s$^{-1}$, which can be provided by
a O9.5-B0 ZAMS star.  The IRAC image shown in the upper panel of Figure
\ref{fig-IR-17238} shows that component A is associated with a deeply
embedded object seen conspicuously in MIPS at 24 \um\ (lower panel), but
undetected in the GLIMPSE images, indicative of very high absorption. Note
that the object located 3\arcsec\ to the west of component A, seen at 4.5
and 3.6 \um\ (GLIMPSE 352.5170$-$00.1540), and in 2MASS bands
(2MASS17271105$-$3519310), is a foreground object. Assuming that the
emission detected in the IRAS bands (IRAS 17238$-$3516) corresponds to
reprocessed emission from the exciting source of component A and that it is
located at its kinematical (near) distance of 5.2 kpc, we derive a total
luminosity of $\sim10^5$~\Lsun, equivalent to that of an O7 ZAMS star. The
ionizing photon rate expected from a star of that luminosity ($N_\star
\sim4.2\times10^{48}$ s$^{-1}$) is about 13 times greater than that derived
from the fit to the radio spectra, indicating that this source is
intrinsically underluminous in radio. Water maser emission has been
detected towards the central radio source \citep{Forster1999A&AS}. Methanol
maser activity has been detected toward this field
\citep{Pestalozzi2005AA}, but its position is not well constrained by the
observations.

We note that the flux at 8.64 GHz reported by \citet{Walsh1998MNRAS} is
$\sim6$ times smaller than the flux measured by us at 8.6 GHz.  Moderate flux
variations in hypercompact \hii\ regions have been detected
\citep[e.g.][]{Galvan-Madrid2008ApJ}, but according to recent models, it is
very unlikely to explain the large  flux increment implied by our
observations in a $\sim10$ year basis \citep{Galvan-Madrid2011MNRAS}. Either the
reason of the difference being observational or physical, we do not discuss
it further in the present work.

Component B corresponds probably to a more developed \hii\ region, which is
consistent with the radio emission at the lower frequencies.  Figure
\ref{fig-IR-17238} shows that this source is associated with a faint 24
\um\ counterpart.

{\it G009.9937$-$00.0299.}-- This jet candidate, taken from
\citet{Walsh1998MNRAS}, is associated with the MSX source
G009.9983$-$00.0334 and with IRAS 18048$-$2019.  It is the only radio
source within the $80\arcsec\times80\arcsec$ region shown in
Fig.~\ref{fig-18048}. Its radio continuum spectrum (see
Fig.~\ref{fig-especjetcands}) can be well modeled with the theoretical
spectra of an homogeneous constant density \hii\ region.  Assuming an
electron temperature of $8000$ K, we derive an emission measure of
$5.9\times10^5$ pc cm$^{-6}$ and an angular size of 1.5\arcsec. The rate of
ionizing photons required to excite the \hii\ region is $7.9\times10^{45}$
s$^{-1}$.  Assuming that this object is located at 5 kpc --- the
\emph{near} distance derived from the observed
$v_{LSR}=49$\,\kms\ \citep{Bronfman1996AAS} --- its total luminosity is
1.7$\times10^4$~\Lsun\ \citep{Lackington2011thesis}, in between those of B0
and a B0.5 ZAMS stars, and for which $N_\star=7.9\times10^{46}$ s$^{-1}$
(interpolated from \citealt{Panagia1973AJ}).  The G009.9937$-$00.0299
object is then underluminous in radio by a factor of $\sim10$.  However, we
do not see neither the spectral nor morphological features to characterize
this radio source as a jet.  Moreover, the three-color MIR image presented in
Fig.~\ref{fig-IR-18048} resembles that of G317.4298$-$00.5612, exhibiting a
single central point source with color characteristic of photoionized
gas. We conclude that this radio source corresponds to an UC \hii\ region.

\citet{Walsh1998MNRAS} reported  methanol
maser emission located approximately 45\arcsec\ toward the southwest 
of the radio source and probably not directly related to it.
Water maser activity has also been detected toward this source
\citep{Valdettaro2001AA}.

{\section{DISCUSSION}\label{sec-discussion}}

\subsection{Nature of the jet candidates}

In this section we summarize the nature of 7 out of the  33 jet candidates
listed in Table 1.  We find that two candidates --- G337.4032$-$00.4037 A
and G345.4938$+$01.4677 --- can be identified, from their radio continuum
spectral characteristics, as ionized jets.  Even though the angular
resolution of the observations was insufficient to resolve the jet
morphology, there is good evidence, such as the presence of collimated
lobes of radio emission and energetic molecular outflows to consider these
two objects as bona-fide collimated ionized  winds.

To determine the physical parameters of the jet found toward
G345.4938$+$01.4677 we assumed a pressure confined jet model
\citep{Guzman2010ApJ}.  This model reproduces well the observed spectral
indexes of G345.4938$+$01.4677 and G337.4032$-$00.4037 A.  Despite of the
fact that 
pressure has been rejected as the collimation mechanism of jets associated
to low-mass stars \citep{Cabrit2007LNP}, it is still a plausible
mechanism in the high-mass case.  Jets associated to HMYSOs 
are found near the center of 
high-mass star forming cores, which  have enough density and temperature to confine
a jet with the characteristics described above. Furthermore, ram pressure
from infalling gas is also relevant in these cores. 
Infalling motions have been detected
toward all cores associated with luminous HMYSOs with jets:
G343.1262$-$00.0620, \citealt{Garay2003ApJ};
G345.4938$+$01.4677, \citealt{Guzman2011ApJ}; 
and G337.4032$-$00.4037 A \citep{Guzman2011thesis}.

Three of the jet candidates, G317.4298$-$00.5612 A, G345.0061$+$01.7944 B, and
G352.5173$-$00.1549 A, turned out to correspond to small ($D < 0.03$ pc) regions of ionized
gas with large emission measures (${\rm EM} > 10^8$ pc cm$^{-6}$), and thus
can be classified as hypercompact \hii\ regions, according to the
definition of \citet{Sewilo2011ApJS}.  HC\hii\ regions are thought to trace
the earliest stages of photoionized nebulae associated to 
single or  binary high-mass star system
system. To date, about a dozen of these regions are known
\citep{Hoare2007prpl}.  Finally, the remaining two jet candidates observed,
G009.9937$-$00.0299 and G305.7984$-$00.2416 A, correspond to UC\hii\ regions
(EM$>10^6$ pc cm$^{-6}$, size $<0.1$ pc).  
An important issue concerning  the
jet list candidates is whether the radio
and FIR emission arise from the same object. 
At least in one
case, G305.7984$-$00.2416 A, we found that the cataloged RMS radio source
is not associated with the luminous object seen in IRAS bands, and that the
much fainter component B is likely to be the youngest and more embedded
source of this high-mass star maternity.

Considering the 7 jet candidates observed as part of this search, plus
G343.1262$-$00.0620, which \citet{Garay2003ApJ}
showed that corresponds to a collimated ionized jet, then there are 8
of the 33 jet candidates for which their nature have been determined.  Of
these, 37.5\% corresponds to collimated ionized winds, 37.5\% corresponds
to HC \hii\ regions and 25\% corresponds to UC\hii\ regions. Due to the
homogeneity of the selection criteria used in building the jet candidate
list (Table \ref{tab-candidates}) we expect the same proportion of objects
in the whole list.  Thus, we expect to find about 9 new ionized jets among
the 25 candidates not yet observed.

\subsection{Lifetime of jets in HMYSOs}

Despite of the limited size of the sample of objects observed in this work,
we can analyze the statistical incidence of jets from our observations and
draw some conclusions about the lifetime of the jet phenomena in the
formation process of high-mass stars.  For the statistical analysis we will
only consider the jet candidates chosen from the RMS survey.  This
selection allows us to make comparison with the results of a recently
reported study of HMYSOs and compact \hii\ regions (C\hii Rs) identified in
the RMS survey \citep{Mottram2011ApJL}. In particular, they estimated that
the lifetime of the C\hii R phase, for all range of
luminosities, is $\sim3\times10^5$ yr.  Of the 239 radio sources we
initially considered from the entire RMS radio sample reported by
\citet{Urquhart2007AA}, we find that 92 are located at angular distances
smaller than 25\arcsec\ from an IRAS point source with a luminosity
$L>2\times10^4$~\Lsun. We assume that these 92 sources form an unbiased
sample of the C\hii\ region population analyzed in \citet{Mottram2011ApJL}.

Of these 92 sources we find that 23 fulfill the two additional criteria
required to be considered jet candidates: positive spectral index and radio
underluminosity by a factor of at least 10. These are the objects with
entry (1) in col.~(10) of Table \ref{tab-candidates}. Since we can not
disentangle the FIR luminosity from the jet candidates G301.1364$-$00.2249 A
and B, and G317.8908$-$00.0578 A and B, we count them as if they were
single objects. At the high luminosity of the sources we are sampling, completeness
corrections of the RMS survey are small: about $0.8$ and $0.97$ for sources
of $20,000$ and $30,000$~\Lsun, respectively, and essentially complete for
more luminous HMYSOs \citep{Mottram2011ApJL}.

We observed four jet candidates drawn from the RMS subsample and found that
two of them are bona-fide jets (G337.4032$-$00.4037 A and
G345.4938$+$01.4677). Using this detection rate, we then expect to find
about 11 jets within the 23 candidates.  The jet incidence in the
population of C\hii\ regions is therefore $\sim$11/92, which in combination
with the C\hii\ region average lifetime of $\sim3\times10^5$ yr
\citep{Mottram2011ApJL} implies that the lifetime of the jet phase is
roughly $4\times10^4$ yr. This lifetime is comparable to the
Kelvin-Helmholtz timescale of a HMYSO of $10^4$~\Lsun. Afterwards, the
central protostar rapidly contracts onto the main sequence producing an
HC\hii\ region. This could explain why few jets are observed associated
with HMYSOs.

Similarly, for the HC\hii\ regions we 
derive lifetimes of the order of $4\times10^4$ yr. 
Multiple radio sources in the field is an issue to consider
within the RMS sample, but we found that all companion sources correspond to
more diffuse and evolved \hii\ regions. 

\subsection{The role of jets in high-mass star formation}

Our systematic search have shown that jets occur in the formation process of 
high-mass stars, at least for HMYSOs with luminosities up to 
$1\times10^5~\Lsun$. This result strongly suggests a disk-mediated accretion 
scenario for the formation of high-mass stars.  
The duration of the jet phase is short, $\sim4\times10^4$ yr, explaining why 
they are not commonly observed.

This age estimate allows us to assess the role of the collimated ionized
jets in driving the larger scale molecular outflows observed toward HMYSOs.
The mean mass loss rate estimated from the values of known jets
\citep[see][]{Garay2003ApJ,Guzman2011ApJ}, is
$\sim5\times10^{-6}\dot{\Msun}$ yr$^{-1}$. Assuming that the jet velocities
are $\sim500\,\kms$
\citep{Anglada1996ASPC,Marti1998ApJ,Curiel2006ApJ,Rodriguez2008AJ,Guzman2010ApJ},
then the momentum delivered by the jets is $\sim100$\Msun~\kms.  On the
other hand, the mean masses and gas velocities of the massive molecular
outflows reported by \citet{Beuther2002AA} are 25 \Msun\ and $28\,\kms$,
respectively, implying a mean momentum of $\sim700$\Msun~\kms.  Thus, even
with the assumption that all of the jet momentum is transferred to the
outflow \citep{Richer2000PrPl}, the jet momentum as estimated above falls
short, by a factor of $\sim7$, to drive the molecular outflow. This factor
is similar to the one reported for G345.4938+01.4677 by \citet{Guzman2011ApJ}.
There are at least two possible answers to explain this discrepancy: i) the
jet is not entirely ionized, but their ionization fraction is about a 10\%
or below; ii) as appears in the radio images, the ejection of mass from the
jet is episodic and highly discontinuous, most of the momentum being
delivered during short bursts, and what we observe now correspond to a more
``quiescent'' phase of the jet.

{\section{SUMMARY}\label{section-summary}}

We are currently undertaking a systematic search of collimated jets towards
high-mass young stellar objects with the goal of determining the pertinence
of this phenomenon in the formation process of massive stars.  In this
paper we present the first results of multi-frequency radio continuum
observations, made using ATCA, towards 6 HMYSOs candidates to harbor jets.
Our main results and conclusions are summarized as follows.

\begin{enumerate}
\item{We compiled a list of 33 HMYSOs likely to contain jets using a selection 
criteria based on their radio and infrared properties. The objects have large 
bolometric luminosities ($L>2\times10^4~\Lsun$) and are underluminous 
at radio wavelengths compared to what is expected from the total luminosity.}

\item{We obtained radio continuum observations at 1.4, 2.4, 4.8, and
  8.6 GHz using ATCA, towards six HMYSOs in the list. We find that one
  object corresponds to an ionized jet, three to hypercompact
  \hii\ regions and two to ultracompact \hii\ regions.}

\item{The jets discovered as part of this search, G337.4032$-$00.4037  and 
G345.4938$+$01.4677 -- the latter reported and analyzed previously by
 \citet{Guzman2010ApJ} -- are associated
  with the two most luminous ($7\times10^4$ and $1.0\times10^5$ \Lsun)
  HMYSOs known to harbor this type of objects. This indicates that the
  phenomena of collimated ionized winds appears in the formation process of
  stars at least up to masses of $\sim 20$ \Msun\ and provides strong
  evidence for a disk-mediated accretion scenario for the formation of
  high-mass stars.  From the rate of occurrence of jets in our sample, we
  estimate that the jet phase in high-mass protostars lasts for
  $\sim4\times10^4$ yr.}

\item{The estimated momentum that the observed ionized jets deliver during
  its main lifetime is about one tenth of the average momentum of the
  molecular outflows associated to HMYSOs.  Whether this is a problem of
  the models or an actual inadequacy of the jets to explain the dynamics of
  molecular outflows remains to be resolved by future observations.}

\item{An important part of the observed candidates were found to correspond
  to hypercompact \hii\ regions, thought to be an early phase in the
  development of the UC\hii R. These hypercompact regions probably appears
  right after the jet phase. We estimated their lifetime in a similar
  manner as done with the jets, obtaining also $\sim40,000$ yr.}

\item{As part of this search, we performed observations  
  toward 5 other HMYSOs that did not satisfied the criteria to be
  jet candidates. In the Appendix we present these observations and 
  physical interpretation of the data.}

\end{enumerate}
\acknowledgments{
A.G. and G.G. gratefully acknowledge support from CONICYT through projects 
FONDAP No. 15010003 and BASAL PFB-06. 
This publication makes use of the Two Micron All Sky Survey database, 
the {\it Spitzer-}GLIMPSE database, the {\it Spitzer-}MIPSGAL database, 
the Red \emph{MSX} Source survey database at 
www.ast.leeds.ac.uk/RMS, and the VizieR catalog access tool, CDS, Strasbourg,
France.}

\appendix
{\section{Other sources of interest}\label{append}}

In this section we report the results of the observations on five radio
sources, chosen from the \citet{Urquhart2007AA} or \citet{Walsh1998MNRAS}
catalogs, but that do not fulfilled all the requirements to be considered
jet candidates.  These sources were selected in an early phase of our
study, and they are most likely HMYSOs. These  sources 
were observed together with the jet
candidates,  and the same  reduction and calibration 
steps described in \S~\ref{section-observations} were applied.

Table \ref{tab-other-obsredpar} gives the phase
tracking center, the on-source total integration times, the FWHM axes of
the synthesized beam obtained at each frequency, and the deconvolved map
noise levels for each observed field. 

\begin{figure}
\includegraphics[width=\textwidth]{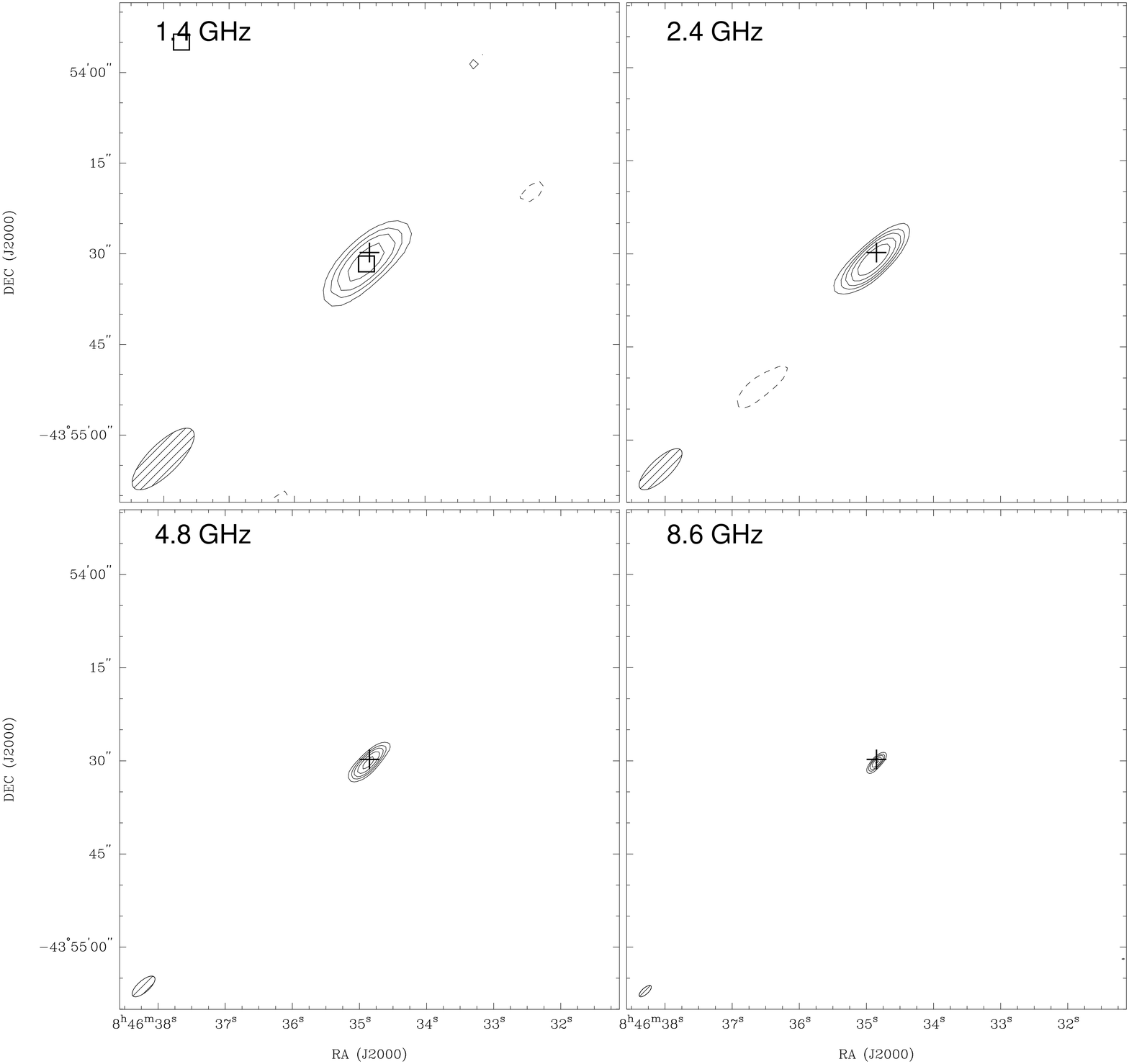} \figcaption {\baselineskip0.0pt ATCA maps of the
  radio continuum emission towards G263.7759$-$00.4281. The cross marks the position
  of the radio source reported by \citet{Urquhart2007AA}. 
  Beams are shown in the lower left corner
  of each panel.  Top left: 1.4 GHz map. The squares marks the peak position 
  of the MSX sources.  Top right: 2.4 GHz map. Bottom left: 4.8 GHz
  map.  Bottom right: 8.6 GHz map.
  Contour levels
  are $-$3, 3, 5, 7, and 10 times 0.15 mJy beam$^{-1}$ for the 1.4 GHz
  image and $-$3, 3, 5, 7, 10, and 14 times 0.15 mJy beam$^{-1}$ for the
  2.4, 4.8, and 8.6 GHz images.
\label{fig-G263}}
\end{figure}

\begin{figure}
\includegraphics[width=\textwidth]{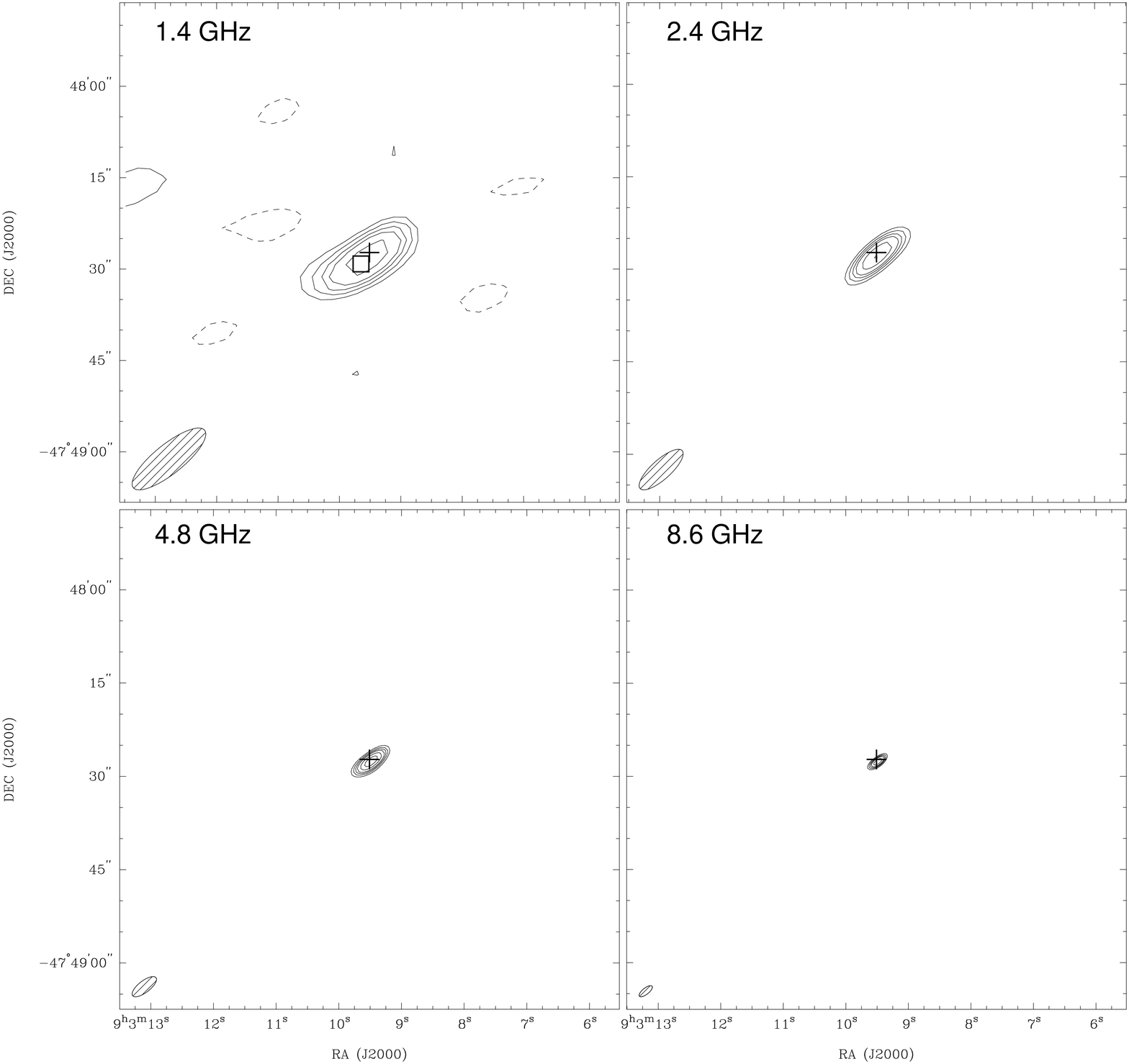} \figcaption {\baselineskip0.0pt ATCA maps of the
  radio continuum emission towards G268.6162$-$00.7389.  The cross marks the position
  of the radio source reported by \citet{Urquhart2007AA}.   Beams are shown in the lower
  left corner of each panel.  Top left: 1.4 GHz map. A square marks the
  position of the MSX source.  Top right: 2.4 GHz map.  Bottom left: 4.8
  GHz map.  Bottom right: 8.6 GHz map. Contour levels
  are $-$5, 5, 7, 9, 11, and 15 times 0.40 mJy beam$^{-1}$ for the 1.4,
  4.8, and 8.6 GHz images, and  $-$5, 5, 7, 9, 11, 15 and 20 times
  0.40 mJy beam$^{-1}$ for the 2.4 GHz image.
\label{fig-G268}}
\end{figure}

\begin{figure}
\includegraphics[width=\textwidth]{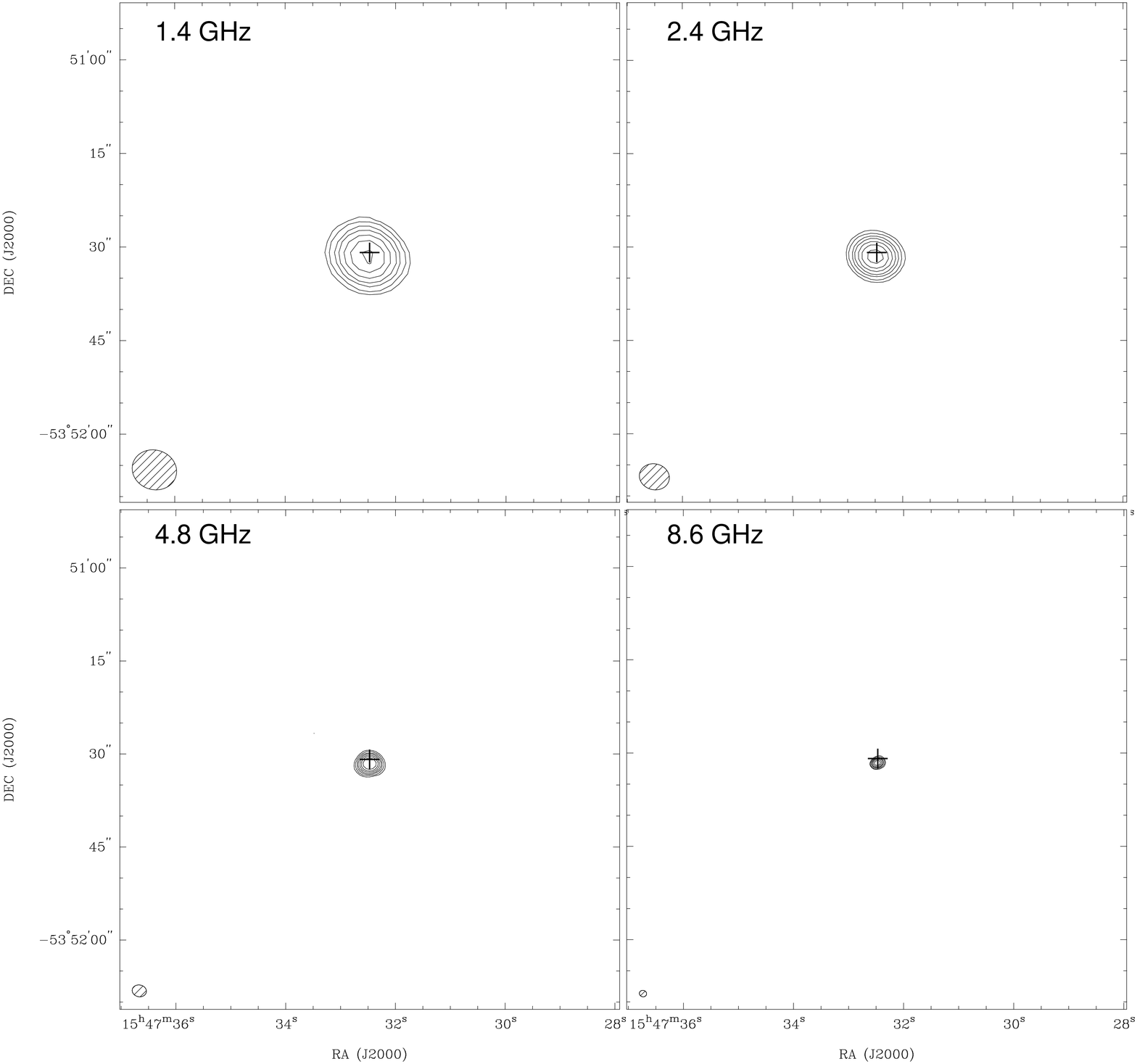} \figcaption {\baselineskip0.0pt ATCA maps of the
  radio continuum emission toward G327.1192+00.5103. The cross marks the position
  of the radio source reported by \citet{Urquhart2007AA}.  Beams are shown
  in the lower left corner of each panel.   Top
  left: 1.4 GHz map.  Top right: 2.4 GHz map.  Bottom left: 4.8 GHz map.
  Bottom right: 8.6 GHz map. Contour levels are $-$5, 5, 8,
  13, 18, 26, 36, 50 times 0.35 mJy beam$^{-1}$ for the 1.4 and 2.4 GHz
  images, and 0.2 mJy beam$^{-1}$ for the 4.8 and 8.6 GHz images.
\label{fig-G327}}
\end{figure}

\begin{figure}
\includegraphics[width=\textwidth]{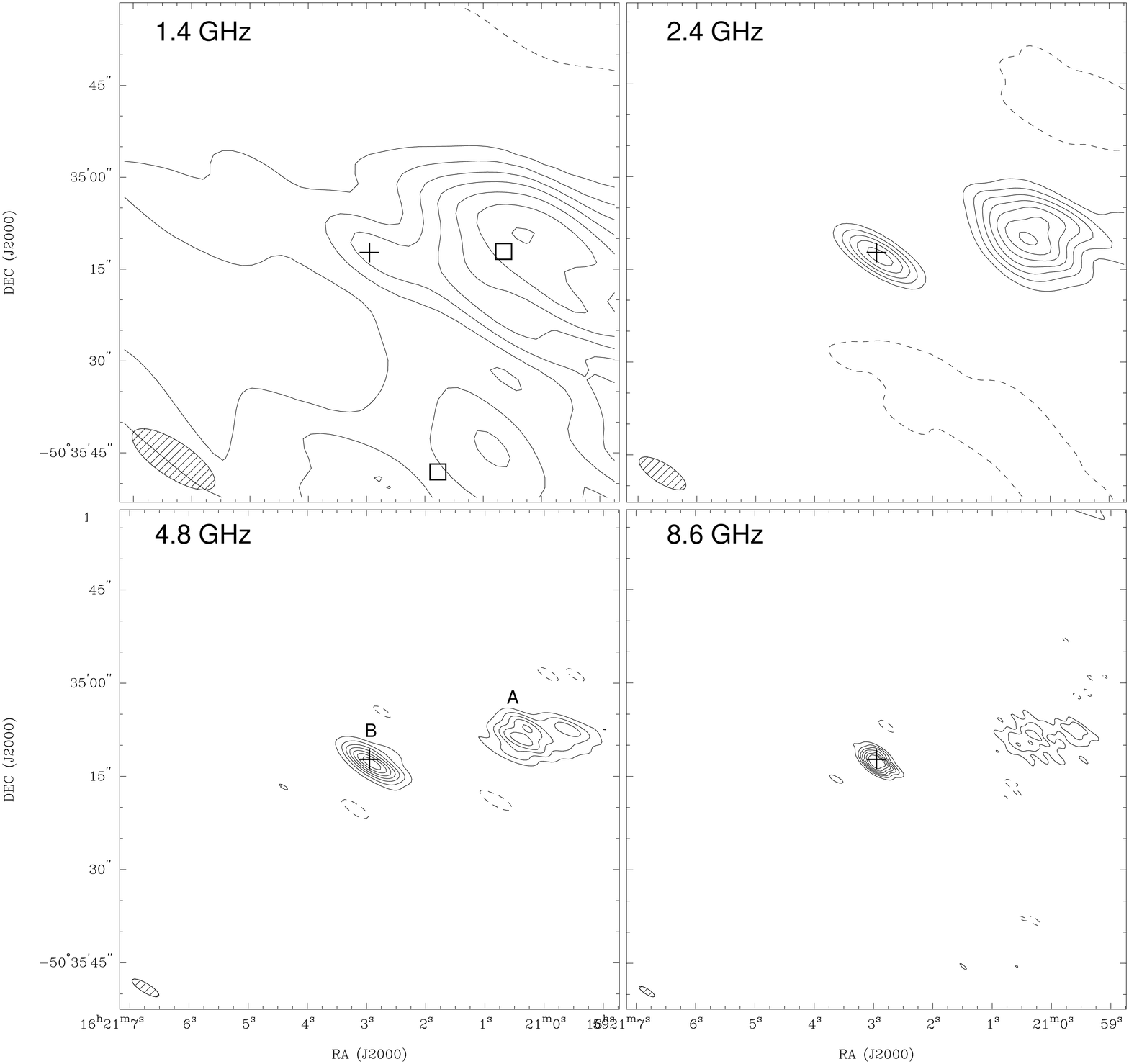} \figcaption
                {\baselineskip0.0pt ATCA maps of the radio continuum
                  emission toward G333.1306$-$00.4275.  The cross marks the
                  position of the radio source reported by
                  \citet{Urquhart2007AA}.  Beams are shown in the lower
                  left corner of each panel.  Top left: 1.4 GHz
                  map. Squares mark the positions of MSX sources.  Top
                  right: 2.4 GHz map.  Bottom left: 4.8 GHz map. The two
                  main radio components are marked with letters A and B.
                  Bottom right: 8.6 GHz map.  Contour levels are $-$7, 7,
                  17, 31, 49, 74, 107, and 150 times 3 mJy beam$^{-1}$.
\label{fig-G333}}
\end{figure}

\begin{figure}
\includegraphics[width=\textwidth]{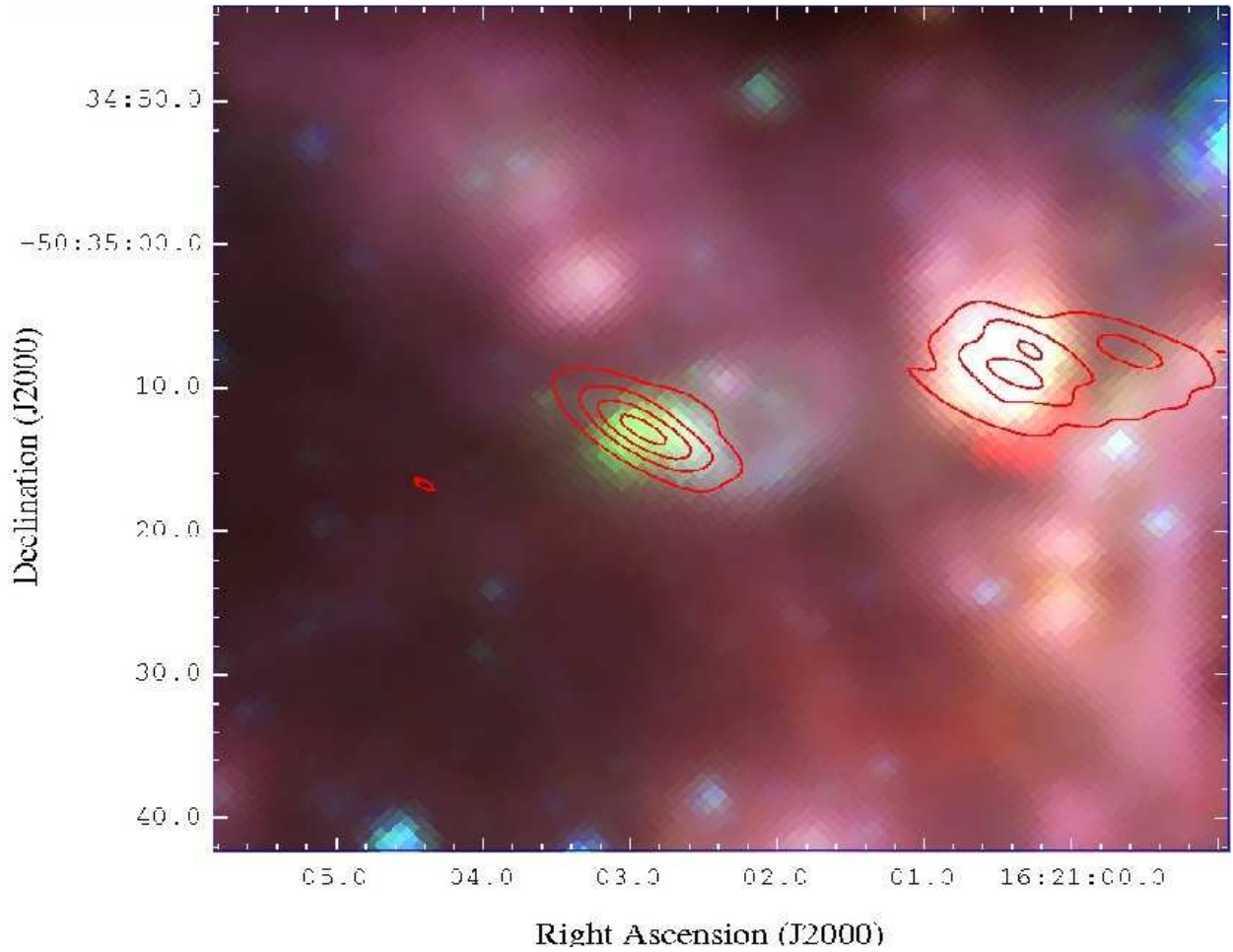} \figcaption
                {\baselineskip0.0pt Three color IRAC image of
                  G333.1306$-$00.4275 made using 8.0, 4.5, and 3.6
                  \um\ data from {\it Spitzer-}GLIMPSE for red, green, and blue,
                  respectively. Intercalated contours of the 4.8 GHz data
                  (Fig.~\ref{fig-G333}) are shown in red.
\label{fig-IR-G333}}
\end{figure}

\begin{figure}
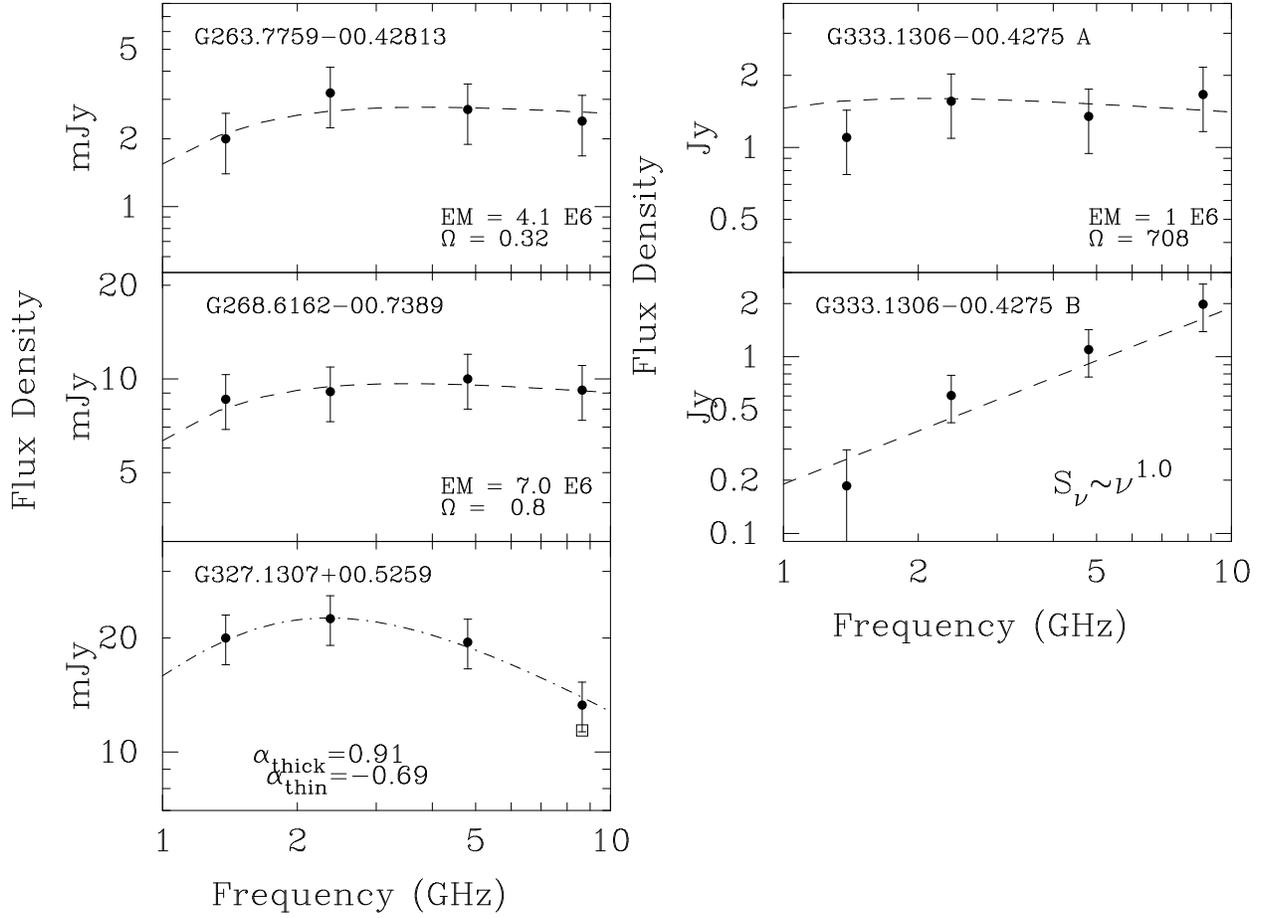

\centering\includegraphics[angle=-90,width=.5\textwidth]{fig22-a.eps}\includegraphics[angle=-90,width=.5\textwidth]{fig22-b.eps}
\figcaption{\baselineskip0.0pt \emph{Left column:} Spectral energy
  distributions of G263.7759$-$00.4281, G268.6162$-$00.7389, and
  G327.1192+00.5103.  In the first and second panel the dashed line shows
  the best-fit obtained with an homogeneous \hii\ region, with the emission
  measure and solid angle indicated in each panel, in units of pc cm$^{-6}$
  and arcsec$^2$, respectively.  In the third panel the dot-dashed line is
  the best fit obtained with a gigahertz peaked source model
  \citep{Snellen1998AAS}.  \emph{Right column:} SEDs of G333.1306$-$00.4275
  A and B.  In the second panel (component B), the dashed line shows the
  best fit obtained with a rising power-law.
\label{fig-spec-other}}
\end{figure}

Emission was detected toward four of the five objects: three of them
correspond to ionized regions associated to a HMYSO, and one to an
extragalactic source.  Maps of the radio continuum emission at the
frequencies of 1.4, 2.4, 4.8, and 8.6 GHz, in regions of
80\arcsec$\times80$\arcsec\ in size, are shown in Figures~\ref{fig-G263}
through \ref{fig-G333}.  The position of the target radio object, as
reported in either catalog of is indicated in the maps with a cross.  We
also show in Figure \ref{fig-IR-G333} a mid-infrared 3 color image taken
toward G333.1306$-$00.4275, that will be used to support its physical
interpretation.  Spectra of the four detected sources are shown in
Fig.~\ref{fig-spec-other}.  The derived parameters, obtained from model
fits to the spectra, are presented in Table \ref{tab-other-pp}. The columns
display the same parameters as Table \ref{tab-pp}, described in
\S~\ref{section-results}.

{\it G263.7759$-$00.42813.}--
This radio source is associated with the RMS/MSX object G263.7759$-$00.4281 
(also IRAS 08448$-$4343), which is known to be associated with an H$_2$O maser 
\citep{Urquhart2009AA} and a H$_2$ bipolar flow \citep{Giannini2005AA}.
The luminosity derived from the IRAS bands and assuming this source is located 
at a distance of 1.6 kpc \citep{Urquhart2007AA13CO}, 
is $\sim 3\times10^3$~\Lsun, therefore failing our luminosity selection criterion. 
It is the only radio source within the $80\arcsec\times80\arcsec$
region shown in Fig.~\ref{fig-G263}. Its radio continuum spectrum 
(see Figure~\ref{fig-spec-other}) can be well modeled with the theoretical 
spectra of an homogeneous constant density \hii\ region.  Assuming an electron 
temperature of $8000$ K, we derive an emission measure of $4.1\times10^6$ pc 
cm$^{-6}$ and an angular size of 0.53\arcsec. The rate of ionizing 
photons required to excite the \hii\ region is $7.0\times10^{44}$ s$^{-1}$. 

Using MSX, IRAS and TIMMI2 data, \citet{Mottram2011AA} derived a bolometric
flux of $F_{\rm bol}\approx 5.77\times10^{-8}$ erg cm$^{-2}$ s$^{-1}$. At a
distance of 1.6 kpc, we derive a bolometric luminosity of $L\sim 4.6\times
10^3$~\Lsun, equivalent to that of B1-B2 zero-age main sequence (ZAMS)
star. The ionizing flux from such a star is $1.4\times10^{45}$ s$^{-1}$
\citep[interpolated from][]{Panagia1973AJ}, similar to that required to
excite the region of ionized gas.  We conclude that this radio object
corresponds to a small (0.004 pc in diameter), optically thin \hii\ region
excited by a single B1-B2 ZAMS star.

{\it G268.6162$-$00.7389.}-- This radio source is associated with the RMS/MSX
source G268.6162$-$00.7389 (also IRAS 09014$-$4736). It is the only radio
source within the $80\arcsec\times80\arcsec$ region shown in
Fig.~\ref{fig-G268}. No water masers were detected toward this source by \citet{Urquhart2009AA} 
at the 0.25 Jy sigma level.
The luminosity derived from the IRAS bands and assuming this source is located 
at a distance of 1.7 kpc \citep{Beck1991ApJ}, 
is $\sim 5\times10^3$\Lsun, therefore failing our luminosity selection criterion. 
 The radio source is unresolved and its radio
continuum spectrum (see Figure~\ref{fig-spec-other}) is well fitted by that of
a uniform density \hii\ region with an emission measure of $7.0\times10^6$
pc cm$^{-6}$, temperature of $15000$ K and an angular size of 0.84\arcsec. 
The ionizing photon flux needed to excite the \hii\ region is
$3.4\times10^{45}$ photons s$^{-1}$.

\citet{Mottram2011AA} reported for this object a bolometric flux of
$7.1\times 10^{-11}$ W m$^{-2}$.  At a distance of
1.7 kpc, the implied bolometric luminosity is
$6.4\times10^3$~\Lsun, equivalent to that of a ZAMS star with a spectral
type between B0.5 and B1.  The ionizing flux from such a star is
$3.4\times10^{45}$ s$^{-1}$ \citep[interpolated from][]{Panagia1973AJ}, 
equal to that required to excite the region of ionized gas. We conclude that 
this radio object corresponds to a small (0.007 pc in diameter), optically thin
\hii\ region excited by a single B0.5-B1 ZAMS star.

{\it G327.1307+00.5259}.-- This radio source is in the field of the RMS
source MSX-G327.1192$+$00.5103, although it is located at an angular
distance of $\sim1$\arcmin\ to the north of the peak MSX emission and the
closest IRAS point source (IRAS 15437$-$5343).  No infrared counterpart has
been detected at MIR wavelengths, neither in {\it Spitzer} (GLIMPSE or
MIPSGAL) nor MSX, which makes it unlikely that this object corresponds to
an embedded HMYSO.

It is the only radio source detected within the $80\arcsec\times80\arcsec$
region shown in Fig.~\ref{fig-G327}. Its radio continuum spectrum (see
Figure~\ref{fig-spec-other}) shows a decrease in the flux density at the higher
frequencies.  Being unresolved at all wavebands, the drop is not produced
by the resolving power of the interferometer.  We suggest that the radio
emission from this source has a non-thermal origin.  From the available
evidence, we further suggest that this is an extragalactic source, possible
a radio-loud galaxy.  Figure \ref{fig-spec-other} shows that the spectrum can
be well fitted with a simple model of a gigahertz peaked galaxy
\citep{Snellen1998AAS}, the spectral indeces in the optically thick and
thin regimes being 0.91 and $-$0.69, respectively.  The lack of infrared
emission could be explained by the source being intrinsically faint at
these wavelengths compared to the radio emission, and/or due to absorption
in the Galactic plane. Similar objects without MIR counterpart had already
been detected in extragalactic surveys \citep{Norris2006AJ,Huynh2010ApJ}.

{\it G333.1306$-$00.4275.}-- This radio object was selected from the
\citeauthor{Urquhart2007AA}~catalog due to its steep spectral index
($\gtrsim$ 1) between 4.8 and 8.6 GHz.  It is located $\sim25$\arcsec\ east
of the closest MSX source G333.1306$-$00.4275 (also IRAS 16172$-$5028),
therefore failing our selection criteria.  There are two radio sources
within the $80\arcsec\times80\arcsec$ region shown in Fig.~\ref{fig-G333}:
a compact central component (the target object; labeled B) and a complex,
extended component, labeled A, associated with the RMS source.

The radio continuum spectrum of component B (see Figure \ref{fig-spec-other})
is well fitted with a power-law spectrum with a spectral index of $\sim
1$, similar to those of HC \hii\ region \citep{Franco2000ApJ}.
The bolometric flux is 15.4$\times10^{-8}$ erg s$^{-1}$ cm$^{-2}$
\citep{Mottram2011AA}, which at a distance of 3.5 kpc \citep{Faundez2004AA}
implies that its total bolometric luminosity is $5.9\times10^4$~\Lsun.

Figure \ref{fig-IR-G333} shows a three-color image of the MIR emission
towards G333.1306$-$00.4275 made using the 3.6, 4.5, and 8.0 \um\ data
obtained from the Spitzer/GLIMPSE survey, and superimposed in red contours
the radio emission detected in 4.8 GHz. This figure clearly shows that
component B is associated with a green extended object, thought to be
related to regions of shocked gas.  Component A is associated with a bright,
extended MIR source, specially bright at 8.0 \um, associated with PAH
emission in a PDR. The radio spectrum is flat
(Fig.~\ref{fig-spec-other}) and probably correspond to
a more evolved \hii\ region.

{\it G345.3768+01.3926.}-- This source was chosen from the catalog of
\citet{Walsh1998MNRAS}.  It is located more than 1\arcmin\ away from the
nearest IRAS source (IRAS 16561$-$4006).  \citeauthor{Walsh1998MNRAS}
reported a peak flux of 5 mJy beam$^{-1}$ at 8.64 GHz and a flux
density\footnote{The integrated fluxes were published online in VizieR
  \citep{Ochsenbein2000AAS}} of 12.3 mJy.  No sources were detected
towards the radio position in GLIMPSE or MSX, which makes this object an
unlikely YSO candidate.  We do not detect emission towards this radio
object in any of the four frequencies at the $\sim0.1$ mJy beam$^{-1}$
level, indicating that this is a highly variable radio source.  We propose
an extragalactic origin. Such extreme variability in radio-loud galaxies is
rare, but has been observed \citep[e.g.,][]{Barvainis2005ApJ}.

\begin{deluxetable}{lcccccccrc}
\rotate  
\tablewidth{0pt}
\tablecolumns{10} 
  \tabletypesize{\small}
  \tablecaption{Jet candidates\label{tab-candidates}}
 \tablehead{
   \colhead{Source Name} &  \colhead{$\alpha$} & \colhead{$\delta$} & \colhead{F$_{\nu\text{-low}}$\tablenotemark{a}}  & \colhead{F$_{\rm 8.6 GHz}$} & \colhead{S.I.}&  \colhead{D}&\colhead{IRAS }  & \colhead{L$_{\it IRAS}$} &\colhead{Ref.} \\
 \colhead{~}  &\colhead{(J2000)} & \colhead{(J2000)}& \colhead{(mJy)} & \colhead{(mJy)} & \colhead{~} & \colhead{(kpc)} &\colhead{~} &\colhead{($10^4$~\Lsun)}&\colhead{~} 
}
\startdata
 G240.3160$+$00.0714   &  07$^{\rm h}$44$^{\rm m}$52\fs04  &  $-$24\arcdeg07\arcmin42\farcs4  & $<$18 & 12.4 &  u  & 7.3 &  07427$-$2400   & 7.7&  (2)\\
 G274.0649$-$01.1460   &  09 24 42.13  &  $-$52 02 00.8  & 28.9 & 36.1 & 0.44   & 6.9\tablenotemark{n} &  09230$-$5148   & 80.&  (1)  \\
 G289.9446$-$00.8909   &  11 01 09.00  &  $-$60 56 56.3  &$<0.7$ & 12.4 &  $\ge2$ & 10.3 &  10591$-$6040   & 21.5&  (1)  \\
 G293.9633$-$00.9776   &  11 32 36.14  &  $-$62 28 08.3  & 24.8 & 28.9 & 0.3    & 11.1 &  11303$-$6211   & 10.1&  (1)  \\
 G298.2234$-$00.3393   &  12 10 01.16  &  $-$62 49 53.9  & 1350.  & 2200.& 0.96 & 11.3 &  12073$-$6233   & 463.&  (1)  \\
 G300.9674$+$01.1499   &  12 34 53.23  &  $-$61 39 40.1  & 102. & 138. & 0.6    & 4.4 &  12320$-$6122   & 22.6&  (1)  \\
 G301.1364$-$00.2249A  &  12 35 35.13  &  $-$63 02 31.7  & 83.8 & 228. & 2.     & 4.4 &  12326$-$6245   & 32.&  (1)  \\
 G301.1364$-$00.2249B  &  12 35 35.19  &  $-$63 02 24.0  & 126. & 179. & 0.68   & 4.4 &  12326$-$6245   & 32.&  (1)  \\
 G305.7984$-$00.2416   &  13 16 42.62  &  $-$62 58 21.2  & $<$10& 5. &  u  & 3. &  13134$-$6242   & 3.4&  (2)\\
 G308.9176$+$00.1231   &  13 43 01.72  &  $-$62 08 56.1  & 247. & 374. & 0.81   & 5.3 &  13395$-$6153   & 25.6&  (1)  \\
 G309.9196$+$00.4791   &  13 50 41.89  &  $-$61 35 11.5  & 377. & 384. & 0.072  & 5.4 &  13471$-$6120   & 27.&  (2)  \\
 G311.1359$-$00.2372   &  14 02 09.93  &  $-$61 58 37.9  & 3.0   & 3.5  & 0.3    & 14.3 &  13585$-$6144   & 11.7&  (1)  \\
 G317.4298$-$00.5612   &  14 51 37.60  &  $-$60 00 19.4  & 6.6 & 8.2 & 0.42     & 15.0 &  14477$-$5947   & 63.4&  (1)  \\
 G317.8908$-$00.0578A  &  14 53 06.19  &  $-$59 20 56.7  & $<0.7$ & 3.6 &  $\ge2$   & 15.1 &  14492$-$5908   & 19.7&  (1)  \\
 G326.4477$-$00.7485   &  15 49 18.67  &  $-$55 16 52.5 & 4.7&  5.9 & 0.39 & 4.3 & 15454$-$5507 & 2.1 & (1) \\
 G328.5759$-$00.5285B  &  15 59 38.15  &  $-$53 45 27.9  & 326 & 381 & 0.27 & 11.4 & 15557$-$5337 & 368. &  (1)  \\
 G331.4181$-$00.3546   &  16 12 50.24  &  $-$51 43 28.6 & 80.7 & 83.9 &0.07 & 4.1 & 16090$-$5135 & 5.7 &  (1) \\
 G333.0162$+$00.7615   &  16 15 18.70  &  $-$49 48 52.8  & 46.9 & 88.8 & 1.2    & 3.3 &  16115$-$4941   & 6.1&  (1)  \\
 G336.9842$-$00.1835   &  16 36 12.42  &  $-$47 37 58.0  & 18. & 34.3 & 1.3     & 10.8 &  16325$-$4731   & 38. &  (1)  \\
 G337.4032$-$00.4037   &  16 38 50.45  &  $-$47 28 02.7  & 90.3 & 130. & 0.71   & 3.2 &  16351$-$4722   & 10.6&  (1)  \\
 G337.7051$-$00.0575   &  16 38 29.63  &  $-$47 00 35.3  & 76.3 & 171. & 1.6    & 12.2 &  16348$-$4654   & 53.1&  (1)  \\
 G337.8442$-$00.3748   &  16 40 26.67  &  $-$47 07 13.1  & 11.1 & 24.4 & 1.5    & 3.1 &  16367$-$4701   & 3.9&  (1)  \\
 G340.0708$+$00.9267     &  16 43 15.69  & $-$44 35 16.0 & 48.8 & 93.8 &1.1      & 4.9 & 16396$-$4429 & 7.2  &  (1) \\
 G340.2768$-$00.2104     & 16 48 53.30   & $-$45 10 22.3 & $<1$ & 4.7 & $\ge 2$  & 3.6 & 16452$-$4504 & 8.0  & (1)  \\
 G343.1262$-$00.0620   &  16 58 17.21  &  $-$42 52 07.1  &$<$7.3  & 4.62 & u  & 2.9 &  16547$-$4247   & 6.3& (2)\\
 G345.0061$+$01.7944   &  16 56 47.59  &  $-$40 14 25.8  & 127. & 209. & 0.98   & 1.7\tablenotemark{n} &  16533$-$4009   & 5.4&  (1)  \\
 G345.4938$+$01.4677   &  16 59 41.61  &  $-$40 03 43.4  & 4.8 & 12.5 & 1.9     & 1.7 &  16562$-$3959   & 7.0&  (1)  \\
 G352.5173$-$00.1549   &  17 27 11.32  &  $-$35 19 32.8  &  $<$14  & 7.5 &  u  & 5.6\tablenotemark{n} &  17238$-$3516   & 9.9 &  (2)  \\
 G000.3138$-$00.2000   &  17 47 09.66  &  $-$28 46 27.7  & 8.3 & 14.5 & 2.2 & 8.0\tablenotemark{n} &  17439$-$2845   & 48.&  (2)   \\
 G009.9937$-$00.0299   &  18 07 52.84  &  $-$20 18 29.3  &$<$10  & 2.32 &  u  & 5.0\tablenotemark{n} & 18048$-$2019   & 2.1&  (2)  \\
 G010.8403$-$02.5913   &  18 19 12.10  &  $-$20 47 30.7 &  $<$21  & 4.8 &  u  & 1.9\tablenotemark{n} & 18162$-$2048   &2.5 &  (2)  \\
 G024.4673$+$00.4910   &  18 34 08.12  &  $-$07 18 18.2  &  $<$28  & 13 &  u  & 5.9 &  18314$-$0720   & 44.&  (2)  \\
 G025.6469$+$01.0534   &  18 34 20.91  &  $-$05 59 39.3  &  $<$7  & 1.4 &  u  & 3.1 &  18316$-$0602   & 2.9&  (2)  \\

%
%
%
%
%
%
%
\enddata
\tablenotetext{a}{Here $\nu$-low represents 4.8 or 6.7 GHz depending whether the candidates comes from 
the RMS  or from \citet{Walsh1998MNRAS} catalog, respectively.}
\tablenotetext{n}{Distance ambiguity not resolved. Near distance adopted.}
\tablerefs{(1):\citet{Urquhart2007AA}, (2): \citet{Walsh1998MNRAS}.}
\end{deluxetable}

\begin{deluxetable}{lcccccccccrrrr}
  \rotate
\tablecolumns{13}
  \tablewidth{0pt} 
  \tabletypesize{\scriptsize}
  \tablecaption{Observational parameters of the observed jet candidates\label{tab-obsredpar}}
 \tablehead{
    \colhead{~} &  \colhead{} & \colhead{} & \colhead{} &    
    \multicolumn{4}{c}{Synthesized Beam}  &\colhead{} &\multicolumn{4}{c}{Noise (mJy beam$^{-1}$)} \\
   \cline{5-8}\cline{10-13}
    \colhead{Source}& \multicolumn{2}{c}{Phase Tracking Center}  & \colhead{Time}
    & \colhead{1.4GHz} & \colhead{2.4GHz} & \colhead{4.8GHz} & \colhead{8.6GHz}
     &\colhead{}& \colhead{1.4GHz} & \colhead{2.4GHz} & \colhead{4.8GHz} & \colhead{8.6GHz} \\
\cline{2-4}\colhead{}   & \colhead{$\alpha$(J2000)}  & \colhead{$\delta$(J2000)}  
    & \colhead{(mins.)} 
& \colhead{(\arcsec)} & \colhead{(\arcsec)} & \colhead{(\arcsec)} & \colhead{(\arcsec)}
 &\colhead{}& \colhead{(mJy)} & \colhead{(mJy)} & \colhead{(mJy)} & \colhead{(mJy)} 
  }
  \startdata
G305.7984$-$00.2416 &13 16 42.62  &   $-$62 58 21.2 & \phn90  & $\phn6.6\times6.3$ &  $\phn4.4\times3.9$ &  $2.2\times1.7$ & $1.1\times0.9$ && 0.2 & 0.2 & 0.07 & 0.08 \\  
G317.4298$-$00.5612 &14 51 37.60  & $-$60 00 19.8   &    110  &$\phn9.9\times5.4$ &  $\phn6.8\times3.8$ &  $2.8\times1.7$ & $2.1\times1.1$  && 0.2 & 0.2 & 0.1  & 0.08 \\  
G337.4032$-$00.4037 &16 38 50.45  & $-$47 28 02.7   &    240  & $\phn8.1\times6.0$ &  $\phn5.4\times3.9$ &  $2.2\times1.7$ & $1.1\times0.9$ && 0.4 & 0.2 & 0.08 & 0.09 \\  
G345.0061$+$01.7944 &16 56 47.59  &  $-$40 14 25.8  &    180  &$\phn8.7 \times5.9$ &  $\phn5.6\times4.0$ &  $2.6\times2.0$ & $1.4\times1.0$ && 0.3 & 0.3 & 0.1  & 0.2  \\  
G352.5173$-$00.1549 &17 27 11.32  &   $-$35 19 32.8 &    120  &$10.1\times5.6$ &  $\phn6.4\times3.7$ &  $3.1\times2.0$ & $1.7\times1.0$     && 0.2 & 0.2 & 0.2 & 0.1 \\    
G009.9937$-$00.0299 &18 07 52.84  &   $-$20 18 29.3 &    110  &$20.1\times5.1$ &  $12.5\times3.0$ &  $5.9\times1.7$ & $3.2\times0.9$        && 0.2 & 0.1 & 0.06 & 0.05 \\  
\enddata
\vspace*{-3 ex}
\end{deluxetable}

 \begin{deluxetable}{lrrcrrrrcrrrr}
   \rotate
   \tablewidth{0pc}
   \tablecolumns{13}
  \tabletypesize{\scriptsize}
  \tablecaption{Observed parameters of radio sources \label{tab-fluxes}}
 \tablehead{ \colhead{~} &\multicolumn{2}{c}{Coordinates}&\colhead{}&\multicolumn{4}{c}{Flux density} & \colhead{} & \multicolumn{4}{c}{Deconvolved Sizes} \\
    \colhead{~}  &\colhead{R.A.} &\colhead{Dec.}  && \colhead{1.4 GHz}  & \colhead{2.4 GHz} & \colhead{4.8 GHz}  & \colhead{8.6 GHz} &\colhead{}&\colhead{1.4 GHz}  & \colhead{2.4 GHz} & \colhead{4.8 GHz}  & \colhead{8.6 GHz}   
\\
    \colhead{Source}&\colhead{(J2000)}&\colhead{(J2000)}&\colhead{}& \colhead{(mJy)}  & \colhead{(mJy)} & \colhead{(mJy)}  & \colhead{(mJy)} &\colhead{}&\colhead{(\arcsec)}  & \colhead{(\arcsec)} & \colhead{(\arcsec)}  & \colhead{(\arcsec)}                                                 
  }
   \startdata
\cutinhead{\textbf{\small Jet candidates}} 
G305.7984$-$00.2416 A   &13$^{\rm h}$16$^{\rm m}$42\fs63&$-$62\arcdeg58\arcmin20\farcs9 &&  10.4  &  11.2 &  11.2 & 9.3  &&  U    &1.4    &1.4    & 1.4    \\
G317.4298$-$00.5612 A   &14 51 37.63&$-$60 00 20.2 &&\nodata&\nodata&  11.0& 19.7  &&\nodata&\nodata& U     & 0.36   \\
G337.4032$-$00.4037 A   &16 38 50.47&$-$47 28 03.1 &&  27.8  & 56.7 &  110.3& 139.5 && 1.6   &  1.2  &  0.5  &  0.8   \\
G345.0061$+$01.7944 B   &16 59 37.76&$-$40 12 03.8 &&  12.2 & 44.4 & 152.5 &
271.8 &&  1.3  &  U    &  0.7  &  0.7   \\
G352.5173$-$00.1549 A   &17 27 11.32&$-$35 19 32.2 && 4.4  & 14.1 &  40.5 &  66.0 &&  U    &   U   &  0.9  &  0.4   \\
G009.9937$-$00.0299     &18 07 52.82&$-$20 18 28.9 && 3.4   &  3.5  &   3.2
&2.3  && U     & U     &  U    & U      \\
\cutinhead{\textbf{\small Additional sources detected in the fields}} 
G305.7984$-$00.2416 B   &13 16 43.23 &$-$62 58 32.9 &&\nodata&\nodata&  0.7  & 1.0  &&\nodata&\nodata& U     & U      \\
G305.7984$-$00.2416 C   &13 16 43.41 &$-$62 58 28.7 && 1.2   &  2.2  &  1.8  & 2.0  &&  U    & U     & U     & 0.9    \\
G305.7984$-$00.2416 D\tablenotemark{a}   &13 16 44.50 &$-$62 58 54.7 &&  32.6 &  27.6 &  24.2 & 35.0 &&  8.0  & 10    & 10    & 10     \\
G317.4298$-$00.5612 B\tablenotemark{a}   &14 51 38.46 &$-$60 00 24.3 &&  41.6 &  48.9 &  5.7 &\nodata&& 7.9  & 7.1   & 7     &\nodata \\
G337.4032$-$00.4037 B\tablenotemark{a}   &16 38 51.13 &$-$47 28 14.8 &&  5.0  &  5.1 &  6.5  &  4.7 && U     &   U   &  2.6  &$\sim2$ \\
G345.0061$+$01.7944 A\tablenotemark{a}   &16 56 45.97 &$-$40 14 38.1 &&  91.4 & 99.1 &  83.0 &  72   &&  7.1  & 8.1   &$\sim8$& $\sim8$\\
G352.5173$-$00.1549 B\tablenotemark{a}   &17 27 11.91 &$-$35 19 35.5 &&  2.8  &  1.6 &\nodata&\nodata&&  U    &   U   &\nodata&\nodata 
\enddata
\tablenotetext{a}{The peak positions for these sources was obtained from
  the 1.4 GHz images.}
\end{deluxetable}
\begin{deluxetable}{lrcrrcrrl}
   \tablewidth{0pc} \tablecolumns{9} \tabletypesize{\small}
   \tablecaption{Derived parameters of the radio sources \label{tab-pp}}
   \tablehead{ \colhead{~} & \colhead{Dist.} &\colhead{T$_e$}&
     \colhead{EM} & \colhead{$\theta$} & \colhead{N$_i$} & \colhead{D} &
     \colhead{$\langle n_e\rangle$} &\colhead{Type}\\ 
\colhead{Source} &\colhead{(kpc)} & \colhead{($10^4$K)} &\colhead{(pc cm$^{-6}$)}
     &\colhead{(\arcsec)} & \colhead{($10^{45}$s$^{-1}$)} & \colhead{(pc)} &
     \colhead{(cm$^{-3}$)} & \colhead{}
}
\startdata
\cutinhead{\textbf{\small Jet candidates}}
G305.7984$-$00.2416 A   & 3.0 &0.8& $2.30\times10^6$& 1.40 & $9.56$
& 0.020  & $1.1\times10^4$ & UC \hii \\ 
G317.4298$-$00.5612 A   &15. &0.8& $1.90\times10^8$& 0.27 & $718$
& 0.020  & $9.8\times10^4$ & HC \hii \\ 
G337.4032$-$00.4037 A   & 3.2 &0.8& $>2\times10^8$&$<0.64$& \nodata &$<0.01$&
$>1.5\times10^5$  & jet  \\ 
G345.0061$+$01.7944 B   & 1.7 &1.5& $4.60\times10^8$& 0.73 & $170$
& 0.006  & $2.8\times10^5$ & HC \hii \\ 
G352.5173$-$00.1549 A   & 5.2 &1.5& $2.80\times10^8$& 0.42 & $318$
& 0.011  & $1.0\times10^5$ & HC \hii \\ 
G009.9937$-$00.0299     & 5.0 &0.8& $5.94\times10^5$& 1.49 & $7.89$
& 0.036  & $4.1\times10^3$ & UC \hii \\ 
\cutinhead{\textbf{{\small Additional sources detected in the fields}}} 
G305.7984$-$00.2416 B   & 3.0 &1.0& $1.3\times10^8$ &  0.066 &
$1.23$ &0.001 & $3.67\times10^5$ & HC \hii  \\ 
G305.7984$-$00.2416 C   & 3.0 &0.8& $1.0\times10^6$ &  0.901 & $1.74$ &0.013 & $8.74\times10^3$ & UC \hii\\ 
G305.7984$-$00.2416 D   & 3.0 &0.8& $2.0\times10^5$ &  8.027 & $27.6$ &0.117 & $1.31\times10^3$ & C \hii \\ 
G317.4298$-$00.5612 B   &15.0 &0.8& $3.5\times10^5$ &  7.030 & $926.$ &0.511 & $8.30\times10^2$ & C \hii \\ 
G337.4032$-$00.4037 B   & 3.2 &0.8& $3.0\times10^5$ &  2.657 &
$5.16$ &0.041 & $2.70\times10^3$ & UC \hii \\ 
G345.0061$+$01.7944 A   & 1.7 &0.8& $5.0\times10^5$ &  8.027 & $22.2$ &0.066 & $2.75\times10^3$ & UC \hii \\ 
G352.5173$-$00.1549 B   & 5.2 &0.8& $5.0\times10^5$ &  1.329 & $5.68$ &0.033 & $3.86\times10^3$ & UC \hii  
\enddata
\end{deluxetable}
\begin{deluxetable}{lcccccccccrrrr}
  \rotate
\tablecolumns{13}
  \tablewidth{0pt} 
  \tabletypesize{\scriptsize}
  \tablecaption{Observational parameters of other sources of interest\label{tab-other-obsredpar}}
 \tablehead{
    \colhead{~} &  \colhead{} & \colhead{} & \colhead{} &    
    \multicolumn{4}{c}{Synthesized Beam}  &\colhead{} &\multicolumn{4}{c}{Noise (mJy beam$^{-1}$)} \\
   \cline{5-8}\cline{10-13}
    \colhead{Source}& \multicolumn{2}{c}{Phase Tracking Center}  & \colhead{Time}
    & \colhead{1.4GHz} & \colhead{2.4GHz} & \colhead{4.8GHz} & \colhead{8.6GHz}
     &\colhead{}& \colhead{1.4GHz} & \colhead{2.4GHz} & \colhead{4.8GHz} & \colhead{8.6GHz} \\
\cline{2-4}\colhead{}   & \colhead{$\alpha$(J2000)}  & \colhead{$\delta$(J2000)}  
    & \colhead{(mins.)} 
& \colhead{(\arcsec)} & \colhead{(\arcsec)} & \colhead{(\arcsec)} & \colhead{(\arcsec)}
 &\colhead{}& \colhead{(mJy)} & \colhead{(mJy)} & \colhead{(mJy)} & \colhead{(mJy)} 
  }
  \startdata
G263.7759$-$00.4281 &08$^{\rm h}$46$^{\rm m}$34\fs85  & $-$43\arcdeg54\arcmin29\farcs8   & \phn60  &
 $14.0\times5.0$ &  $\phn9.0\times3.3$ &  $4.6\times1.9$ & $2.5\times1.0$    && 0.1 & 0.1 & 0.07 & 0.08 \\  
G268.6162$-$00.7389 &09 03 09.51  & $-$47 48 27.3   & \phn60  &$13.0\times4.4$ 
&  $\phn9.2\times3.2$ &  $4.7\times1.9$ & $2.6\times1.0$     && 0.3 & 0.1 & 0.06 & 0.07 \\  
G327.1307+00.5259 &15 47 32.47 & $-$53 51 30.9 & 180 &$\phn7.2\times6.3$ &
$\phn4.9\times4.2$ & $2.3\times1.9$ & $1.2\times1.0$ && 0.2 & 0.1 & 0.09 &
0.07 \\
G333.1306$-$00.4275 &16 21 02.95  & $-$50 35 12.3   & \phn60  &$12.1\times4.4$ &
  $\phn7.0\times2.7$ &  $4.9\times1.5$ & $2.8\times0.8$     && 3.1 & 1.8 & 2.1  & 3.4  \\  
G345.3768$+$01.3926 &16 59 37.75 & $-$40 12 03.5   &    170
&$\phn9.0\times5.6$ &  $\phn5.7\times3.7$ &  $2.6\times2.0$ &
$1.4\times1.0$  && 0.1 & 0.1 &  0.08 & 0.08\\
\enddata
\vspace*{-3 ex}
\end{deluxetable}

\begin{deluxetable}{lrrcrrrrcrrrr}
   \rotate
   \tablewidth{0pc}
   \tablecolumns{13}
  \tabletypesize{\scriptsize}
  \tablecaption{Observed parameters of other radio sources of interest \label{tab-other-fluxes}}
 \tablehead{ \colhead{~} &\multicolumn{2}{c}{Coordinates}&\colhead{}&\multicolumn{4}{c}{Flux density} & \colhead{} & \multicolumn{4}{c}{Deconvolved Sizes} \\
    \colhead{~}  &\colhead{R.A.} &\colhead{Dec.}  && \colhead{1.4 GHz}  & \colhead{2.4 GHz} & \colhead{4.8 GHz}  & \colhead{8.6 GHz} &\colhead{}&\colhead{1.4 GHz}  & \colhead{2.4 GHz} & \colhead{4.8 GHz}  & \colhead{8.6 GHz}   
\\
    \colhead{Source}&\colhead{(J2000)}&\colhead{(J2000)}&\colhead{}& \colhead{(mJy)}  & \colhead{(mJy)} & \colhead{(mJy)}  & \colhead{(mJy)} &\colhead{}&\colhead{(\arcsec)}  & \colhead{(\arcsec)} & \colhead{(\arcsec)}  & \colhead{(\arcsec)}                                                 
  }
   \startdata
G263.7759$-$00.4281     &08 46 34.848&$-$43 54 30.27 && 2.0   &  3.2  &   2.7 & 2.4  && U     & U     &  U    & U      \\
G268.6162$-$00.7389     &09 03 09.499&$-$47 48 27.66 && 8.6   &  9.1  &   10.0& 9.2  &&  U    & 1.6  &  1.1  & 0.7   \\
G327.1307+00.5259     &15 47 32.472&$-$53 51 31.58 && 20.0  &  22.5 &  19.5 & 13.3 &&   U   & 0.9   &  U    & 0.36   \\
G333.1306$-$00.4275 B   &16 21 02.921&$-$50 35 12.92 && 186   & 604  & 1100  & 2000  &&  U    &  U    &  5.9  &  3.3   \\
G345.3768+01.3926\tablenotemark{*}     &\multicolumn{2}{c}{\nodata}&&\multicolumn{4}{c}{\nodata}&&\multicolumn{4}{c}{\nodata}\\
\cutinhead{\textbf{\small Additional sources detected in the fields}} 
G333.1306$-$00.4275 A   &16 21 00.30\phn &$-$50 35 08.30 && 1100  &1589  & 1348  &1664   && \multicolumn{4}{c}{$\sim25$}\\
\enddata
\tablenotetext{*}{This source was not detected in our observations.} 
\end{deluxetable}

\begin{deluxetable}{lrcrrcrrl}
   \tablewidth{0pc} \tablecolumns{9} \tabletypesize{\small}
   \tablecaption{Derived parameters of the radio sources \label{tab-other-pp}}
   \tablehead{ \colhead{~} & \colhead{Dist.} &\colhead{T$_e$}&
     \colhead{EM} & \colhead{$\theta$} & \colhead{N$_i$} & \colhead{D} &
     \colhead{$\langle n_e\rangle$} &\colhead{Type}\\ 
\colhead{Source} &\colhead{(kpc)} & \colhead{($10^4$K)} &\colhead{(pc cm$^{-6}$)}
     &\colhead{(\arcsec)} & \colhead{($10^{45}$s$^{-1}$)} & \colhead{(pc)} &
     \colhead{(cm$^{-3}$)} & \colhead{}
}
\startdata
\cutinhead{\textbf{\small Other sources of interest}}
G263.7759$-$00.4281     & 1.6 &0.8& $4.05\times10^6$& 0.53 & $0.697$
& 0.004  & $3.2\times10^4$ & UC \hii \\ 
G268.6162$-$00.7389     & 1.7 &1.5& $7.00\times10^6$& 0.84 & $3.40$
& 0.007  & $1.9\times10^4$ & UC \hii \\ 
G333.1306$-$00.4275 B   & 3.5 &1.5& $>2\times10^8$ &$<2.5$ & \nodata   &$<0.042$
&$>7.1\times10^{4}$         & HC \hii \\ 
\cutinhead{\textbf{{\small Additional sources detected in the fields}}} 
G333.1306$-$00.4275 A   & 3.5 &0.8& $1.0\times10^6$ & 24.998 & $1820$ &0.424 & $1.54\times10^3$ & C \hii \\ 
\enddata
\end{deluxetable}
\clearpage


\bibliographystyle{apj}
\bibliography{bibliografia}







\clearpage



\end{document}